\documentclass[]{elsarticle}
\makeatletter
\def\ps@pprintTitle{%
 \let\@oddhead\@empty
 \let\@evenhead\@empty
 \def\@oddfoot{\centerline{\thepage}}%
 \let\@evenfoot\@oddfoot}
\makeatother

\usepackage{graphicx}
\usepackage{amssymb}
\usepackage{array}
\usepackage{subfig}
\usepackage{color}

\newcommand{\omt}[1]{}

\newcommand{\ER}	{Erd\H{o}s-R\'{e}nyi}

\newcommand{\s}		{\ensuremath{s}}		
\newcommand{\sopt}	{\ensuremath{s_{opt}}}		
\newcommand{\W}		{\ensuremath{W}}		
\newcommand{\w}		{\ensuremath{w}}		

\newcommand{\Texp}	{\ensuremath{\overline{F}}}	
\newcommand{\Texpi}[1]	{\ensuremath{\overline{F}(#1)}}	
\newcommand{\Lsexp}	{\ensuremath{\overline{L}_s}}		
\newcommand{\Lexp}	{\ensuremath{\overline{L}}}		
\newcommand{\Lexpopt}	{\ensuremath{\overline{L}_{opt}}}		

\newcommand{\N}		{\ensuremath{N}}		
\newcommand{\kave}      {\ensuremath{\overline{k}}}     
\newcommand{\kRWexp}    {\ensuremath{\overline{k}_{rw}}}
\newcommand{\R}		{\ensuremath{\overline{R}}}		
\newcommand{\pd}	{\ensuremath{d}}		
\newcommand{\pa}	{\ensuremath{a}}		

\newcommand{\pn} 	{\ensuremath{P_n}}
\newcommand{\ps} 	{\ensuremath{P_{\!t\!p}}}
\newcommand{\pf} 	{\ensuremath{P_{\!f\!p}}}
\newcommand{\ptp}       {\ensuremath{p_{t\!p}}}
\newcommand{\pfp}       {\ensuremath{p_{\!f\!p}}}
\newcommand{\PWx}[1]    {\ensuremath{P_{pw}(#1)}}
\newcommand{\pnoderes}  {\ensuremath{p_{res}}}

\newcommand{\nk}        {\ensuremath{n_k}}

\newcommand\blfootnote[1]{%
  \begingroup
  \renewcommand\thefootnote{}\footnote{#1}%
  \addtocounter{footnote}{-1}%
  \endgroup
}

\begin{document}
\begin{frontmatter}


\title{Resource location based on precomputed partial random walks in dynamic networks\blfootnote{Supported in part by Ministerio de Economia y Competitividad grant TEC2014-55713-R, Regional Government of Madrid (CM) grant Cloud4BigData (S2013/ICE-2894, co-funded by FSE \& FEDER), NSF of China grant 61520106005, and European Commission H2020 grants ReCred and NOTRE.}
}


\author{V\'{\i}ctor M. L\'opez Mill\'{a}n \fnref{CEU}}
\author{Vicent Cholvi \fnref{UJI}}
\author{Antonio Fern\'andez Anta \fnref{INN}}
\author{Luis L\'opez \fnref{LADyR}}

\address[CEU]{Universidad CEU San Pablo, Spain}
\address[UJI]{Universitat Jaume~I, Spain}
\address[LADyR]{Universidad Rey Juan Carlos, Spain}
\address[INN]{Institute IMDEA Networks, Spain}


\begin{abstract}
The problem of finding a resource residing in a network node (the \emph{resource location problem}) is a challenge in complex networks due to aspects as network size, unknown network topology, and network dynamics. The problem is especially difficult if no requirements on the resource placement strategy or the network structure are to be imposed, assuming of course that keeping centralized resource information is not feasible or appropriate. Under these conditions, random algorithms are useful to search the network. A possible strategy for static networks, proposed in previous work, uses short random walks precomputed at each network node as partial walks to construct longer random walks with associated resource information. In this work, we adapt the previous mechanisms to dynamic networks, where resource instances may appear in, and disappear from, network nodes, and the nodes themselves may leave and join the network, resembling realistic scenarios. We analyze the resulting resource location mechanisms, providing expressions that accurately predict average search lengths, which are validated using simulation experiments. Reduction of average search lengths compared to simple random walk searches are found to be very large, even in the face of high network volatility. We also study the cost of the mechanisms, focusing on the overhead implied by the periodic recomputation of partial walks to refresh the information on resources, concluding that the proposed mechanisms behave efficiently and robustly in dynamic networks.
\end{abstract}


\end{frontmatter}

\section{Introduction}

\emph{Random walks} are network routing mechanisms which have been extensively studied and used in  a wide range of applications: physics, mathematics,  population dynamics, bioinformatics, etc.~\cite{rw:Hughes95,rw:Lovasz93,rw:Newman06}. Roughly speaking, they choose, at each point of the route, the next node uniformly at random among the neighbors of the current node.

Among the advantages of random walks when applied to communication networks is the fact they need only local information, avoiding the bandwidth overhead necessary in other routing mechanisms to communicate with other nodes. This is especially useful when there is no knowledge on the structure of the whole network, or when the network structure changes frequently. For these reasons, random walks have been proposed as a base mechanism for multiple network applications, including network sampling~\cite{rw:Fortunato06,rw:Lee06}, network resource location~\cite{rw:Adamic01,rw:Gkantsidis06,rw:Rodero10,rw:Yang05}, network construction~\cite{rw:Chawathe03,rw:Law03,rw:Lv02b,rw:Lv02,rw:Mabrouki07}, and network characterization~\cite{rw:Dolev06,rw:Sadagopan05,rw:Tadic01}.

The emergence of the peer-to-peer (P2P) architecture model has been proven useful in many applications in recent years. While structured P2P systems (e.g., Chord~\cite{rod:Stoica01}, CAN~\cite{Ratnasamy:2001:SCN:964723.383072} --Content-Addressable Network--, Kademlia~\cite{Maymounkov:2002:KPI:646334.687801}, etc.) provide efficient search mechanisms, they introduce a significant management overhead. In turn, unstructured systems have little management overhead and, consequently, have been considered in several scenarios (e.g., Gnutella~\cite{rw:Chawathe03,rw:Lv02}, CAP~\cite{Krishnamurthy:2001:EMC:505202.505216} --Cluster-based Architecture for P2P--, etc.). For such systems, searching techniques based on \emph{flooding}, \emph{supernodes} and \emph{random walks} have been used. However, it is known that flooding mechanisms do not scale well~\cite{rod:Jovanovic01}, and supernode systems are vulnerable to supernodes failures (technical problems, attacks, censorship, etc.). Therefore, random walks have been used to search for resources held in the nodes of a network (e.g., \cite{rw:Chawathe03,Manku:2004:KTN:1007352.1007368}), a problem usually known as \emph{resource location}. The problem consists of finding a node that holds a given resource, the \emph{target node}, starting at some \emph{source node}. The source node is checked for the resource: if it is not found locally, the search hops to a random neighbor, checking that node for the resource. The search proceeds through the network in this way, until the target node is reached.

Nevertheless, by using random walks, some nodes may be (unnecessarily) visited more than once, while other nodes may remain unvisited for a long time.   Avoiding this problem is the main objective of our study.

\subsection{The Dynamic Resource Location Problem}

In this work, we are concerned with the resource location problem in networks with dynamic behavior regarding both resources and nodes. 

In particular, we consider scenarios in which resources are randomly placed in the nodes across the network. Then, on the one hand, we consider scenarios in which the instances of the resources may appear and disappear from a time instant to another, maybe at different nodes. On the other hand, we also consider scenarios in which the network nodes themselves may also leave and join the network.

In these scenarios, all the nodes of the network may launch independent searches for different resources (e.g., files) at any time, without the help of a centralized server, and we are interested in measuring the average performance of searches between \emph{any} pair of nodes.

Our assumptions regarding dynamicity cover a wide range of scenarios. For instance, in P2P networks nodes represent users, which may leave and join   the network quite often. Also, resources represent the shared files, which may appear and disappear from time to time.

\subsection{Contributions}

In this work, we use the technique of concatenating \emph{partial random walks} (PWs) to generalize the resource location mechanisms introduced in~\cite{Lopez:Computing} for static networks to the case of dynamic networks. 
In particular, this paper provides new analytical models that predict the behavior of the resource location mechanisms in scenarios with dynamic resources and in scenarios with dynamic nodes, along with new simulation experiments to validate the analytical results. In addition, a new analysis of the cost of the mechanisms in these scenarios is provided.
We consider the two versions of the mechanisms proposed in~\cite{Lopez:Computing} and adapt them to operate in the dynamic scenarios. 
In the first version, which we refer as \emph{choose-first PW-RW}, the search mechanism first chooses one of the PWs at random and then checks its associated information for the desired resource. In the second version, which we refer as \emph{check-first PW-RW}, the search mechanism first checks the associated resource information of all the PWs of the node, and then randomly chooses among the PWs with a positive result. It is clear that there are other choices regarding the search mechanisms that seem reasonable. However, the ones considered in our study follow very different approaches (one chooses first, and the other checks first). Therefore, that will allow us to check the strength of our approach in very different circumstances.

Then, we have studied their performance, considering the following aspects:

\begin{itemize}
\item{\it Dynamic Resources:}
We have developed an analytical mean-field model for both mechanisms when resources are dynamic. Expressions are given for the corresponding \emph{expected search length} (i.e., the expected number of hops taken to find the resource, averaged over all source nodes, target nodes, and network topologies) of each mechanism. These expressions provide predictions as a function of several parameters of the model, such as the network structure (size and degree distribution), the resource dynamics, and those of the mechanisms operation.

The predictions of the models are validated by simulation experiments in three types of randomly built networks: regular, \ER, and scale-free. These experiments are also used to compare the performance of both mechanisms, and to investigate the influence of the resource dynamics. We have  compared the performance of the proposed search mechanisms with respect to random walk searches. For the \emph{choose-first PW-RW} mechanism we have found a reduction in the average search length with respect to simple random walk ranging from around $57\%$ to $88\%$. For the \emph{check-first PW-RW} mechanism such a reduction is even bigger, achieving reductions above $90\%$.

\item{\it Dynamic Nodes:}
We have also considered the case where network nodes may leave and join the network, and have provided both analytical and experimental results. We have found a reduction in the average search length with respect to simple random walks above $94\%$ (using the \emph{check-first PW-RW} mechanism).

\item{\it Cost:}
Finally, we have analyzed the cost of the PW-RW mechanisms, defined as the number of messages, taking into account the cost of searches themselves and the cost of precomputing the PWs in each recomputation interval. We have provided analytical expressions for the relation between the cost and the length of the recomputation interval, as well as for the interval length that minimizes this cost. We have found that the impact of the precomputation of PWs on the cost is not significant in a wide range of lengths of the precomputation interval depending on the dynamic behavior of the network and on the dynamic behavior of searches.

\end{itemize}

\subsection{Related Work}

Da Fontoura Costa and Travieso \cite{rw:Costa07} study the network coverage of three types of random walks: traditional, preferential to untracked edges, and preferential to unvisited nodes. Also, Yang \cite{rw:Yang05} studies the search performance of five random walk variations: no-back (NB), no-triangle-loop (NTL), no-quadrangle-loop (NQL), self-avoiding (SA) and high-degree-preferential self-avoiding (PSA). Self-avoiding walks (SAW) are those that try not to visit nodes that have already been visited. Several variations of this idea have been studied, differing in the probability of revisiting a node. Some examples are: strict SAW, true or myopic SAW, and weakly SAW~\cite{rw:Amit83,rw:Slade09}. In~\cite{rw:DasSarma10}, Das Sarma et al. propose a distributed algorithm to obtain a random walk of a specified length $\ell$ in a number of rounds\footnote{A \emph{round} is a unit of discrete time in which every node is allowed to send a message to one of its neighbors. According to this definition, a simple random walk of length $\ell$ would then take $\ell$ rounds to be computed.} proportional to $\sqrt{\ell}$.

L\'opez Mill\'an et al~\cite{Lopez:Computing} propose a mechanism for resource location based on building random walks connecting together partial walks (PW) previously computed at each network node.  However, the mechanisms in~\cite{Lopez:Computing} are only valid when both nodes and resources have a static behavior, contrary to the approach we follow in this paper. \\ \\

The remainder of this paper is arranged as follows. Sections~\ref{sec:choose_first} and~\ref{sec:check_first} respectively present the \emph{choose-first PW-RW} and \emph{check-first PW-RW} mechanisms in scenarios with dynamic resources. Section~\ref{sec:dyn_nodes} adapts the previous mechanisms to scenarios with dynamic nodes. In Section~\ref{sec:cost}, the cost of the mechanisms is evaluated in dynamic scenarios. Finally, Section~\ref{sec:conclusions} concludes this paper and provides some future work lines.

\section{Choose-First PW-RW with Dynamic Resources}
\label{sec:choose_first}

Consider a communication network (e.g., the Internet, a wireless ad-hoc network, etc.) that provides full connectivity to the end system entities (e.g., computers, smartphones, etc.) connected to it. Next, consider a subset of $N$ end system entities which establish logical neighboring relations for some purpose (a P2P file sharing system, a social network application, etc.). These end system entities (as the \emph{nodes}) and their neighboring relations (as the \emph{links}) form an \emph{overlay network} on top of the \emph{underlying network}.\footnote{Note that neighbors in the overlay are not in general neighbors in the underlying network. Note also that the underlying communication network provides connectivity between \emph{any pair} of end system entities, even if they are not neighbors in the overlay.}
The resource location mechanisms described in this paper apply to such an overlay network (referred to simply as the \emph{network}), and we will focus on searches for resources in the end system entities (referred to as the \emph{nodes}).


Each of the $N$ nodes holds a set of resources. We focus on a given resource of interest, of which initially there is a number of instances randomly placed in many distinct network nodes. Our resource location problem is defined as finding one of the nodes that hold the resource (the one we encounter first, called the \emph{target node}), starting by a certain node (the \emph{source node}). We make no assumption on the underlying communication network, focusing on the number of nodes of the overlay visited to find a resource, as a measure of search performance.

In our analysis, for each search, we assume that the source node is uniformly chosen at random among all nodes in the network. Likewise, we consider that the instances of the resource have been randomly distributed throughout the network. The probability that a given node holds an instance of the resource is denoted by \pnoderes. The expected number of instances of the resource for a given network is denoted by $\R=\N\cdot\pnoderes$.

\begin{figure}
 \centering
 \includegraphics[width=8cm,angle=270]{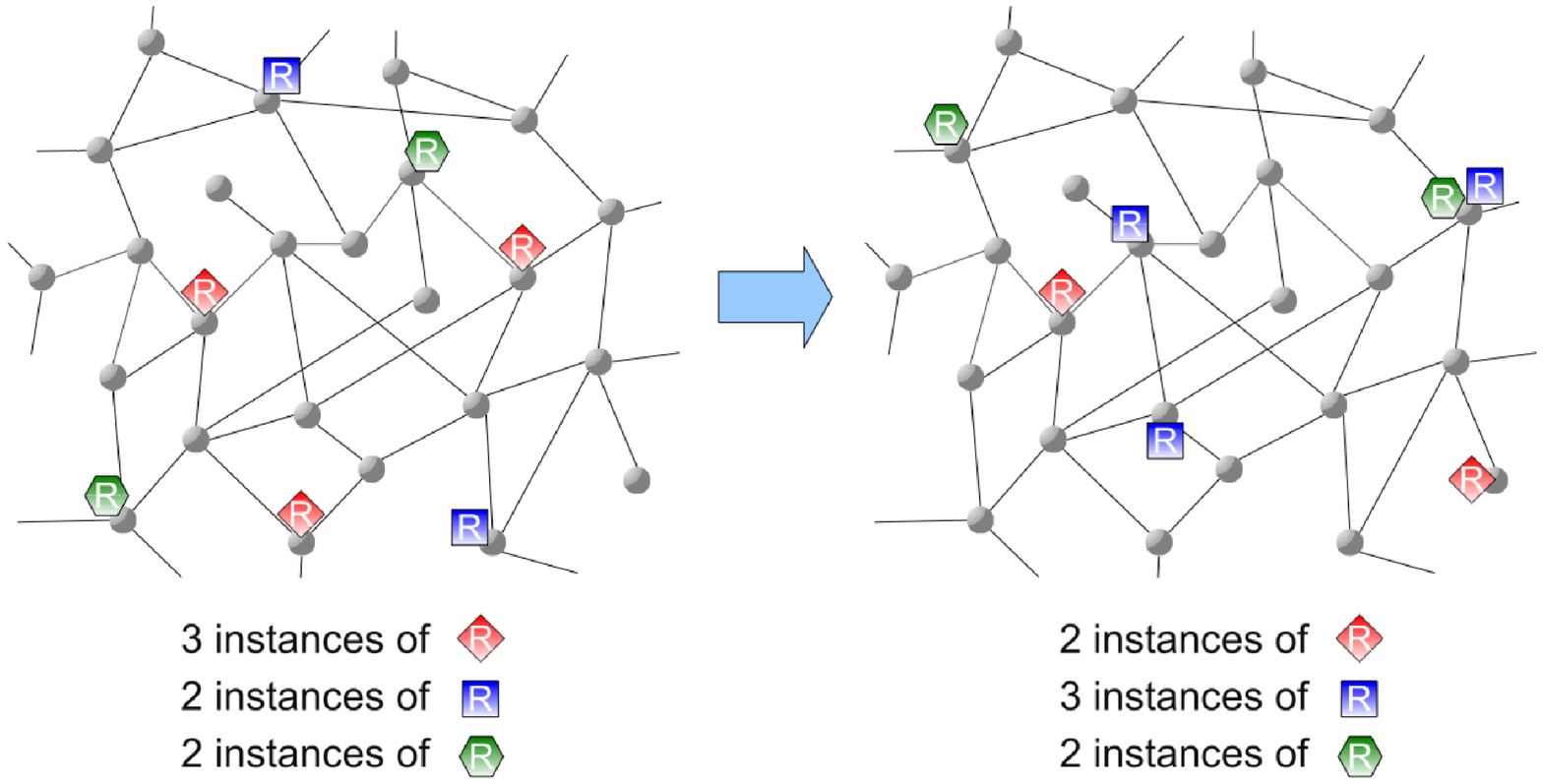}
 \caption{Example of the dynamic behavior of resources. The network is shown at $t\!=\!0$ (left hand side), when the PWs are precomputed, and at time $t=T$ (right hand side), the end of the PW recomputation interval. The resource instances have changed during the interval.} 
 \label{fig:resources}
\end{figure}

Resources have a dynamic behavior. If we compared two snapshots of the network, one taken at time $t\!=\!0$ and another taken at time $t\!=\!T$, we would observe that some of the instances have disappeared while other new ones have appeared (at different nodes, in general).
More concretely, an instance present in a node disappears with probability \pd. Conversely, an instance not initially present in a node appears in that node with probability \pa. We will use \pd\ as an input parameter to characterize resource dynamics. For a value of \pd, we will set \pa\ so that expectation of the number of resources (\R) remains unchanged. This way, our results will isolate the impact of resource dynamics on the search mechanism due only to the deterioration of information, discarding the effect of a possible increase or decrease on the expected number of resource instances. Figure~\ref{fig:resources} provides an illustrative example of the dynamic behavior of the resources.

\paragraph{The search mechanism}
A search performs a walk from the source node to the target node according to the mechanism that is defined below. The search mechanism proposed in this paper, referred to as PW-RW, exploits the idea of efficiently building \emph{total random walks} from \emph{partial random walks} available at each node of the network. It comprises two stages:

\begin{enumerate}
\label{point1}
\item
\emph{Partial random walks construction}: In an initial stage at time $t\!=\!0$, every node $i$ in the network precomputes a set $\W_i$ of \w\ random walks before the searches take place, with the initial distribution of resource instances in the network. Each of these partial walks (PW) has length \s, starting at $i$ and finishing at a node reached after \s\ hops. Using the PW-RW mechanisms, the PWs computed in this stage are simple random walks (i.e., the next node to be visited is chosen uniformly at random among the neighbors of the current node).

During the computation of each PW in $\W_i$, node $i$ registers the resources held by the \s\ first nodes in the PW (from $i$ to the one before the last node). The last node of the PW is excluded, being included in the PWs departing from it. 
In particular, for each PW computed by $i$, this node keeps the set of the identifiers of the resources held by the nodes in that PW. In this set, there is no indication of the particular node or nodes holding the resource.
The registered information will be used by the searches in stage~2,
to decide whether to traverse that PW (if the resource looked for is found in the set of identifiers), or to jump over it (if the resource is not found).

\item
\label{point2}
\emph{The searches}: During the interval $0 < t \leq T$, after the PWs are constructed, searches are performed in the network. We will consider the system at $t\!=\!T$, in which, as stated above, the dynamic behavior of resource instances is characterized by \pd. Therefore, results obtained will reflect the performance of the search mechanism in a worst case scenario, since searches executed in $t < T$ will see a probability that an instance disappears less than or equal to \pd.
There is no relation between the interval $T$, measured in time units, and the hops of a search, other than the assumption that $T$ is much longer than the duration of a typical search, as discussed later in this section in paragraph \emph{Resource dynamics}. 

A search can be qualitatively described as a sequence of jumps over PWs, interleaved with some occassional unsuccessful PW traversals, and finished by the successful traversal of a PW until the target node is visited. 
Unsuccessful PW traversals are caused by outdated resource information associated to that PW, i.e., the resource was in the PW at $t=0$ but it has disappeared at the time of the search.
The last PW traversal will be incomplete in general, in the sense that its length will be less than or equal to \s, since the search stops when the resource is found. 

We measure the length of searches in \emph{hops}. Some of these hops are \emph{jumps} (over PWs), and other are \emph{steps} (traversing PWs). We distinguish between \emph{unnecessary steps} (in unsuccessful PW traversals), and \emph{final steps} (in the last, successful, PW).
The definition of the search mechanism and the associated concepts are illustrated by the example in Figure~\ref{fig:concepts}, in which PWs of length $\s=6$ are used.

\begin{figure}
 \centering
 \includegraphics[width=8cm]{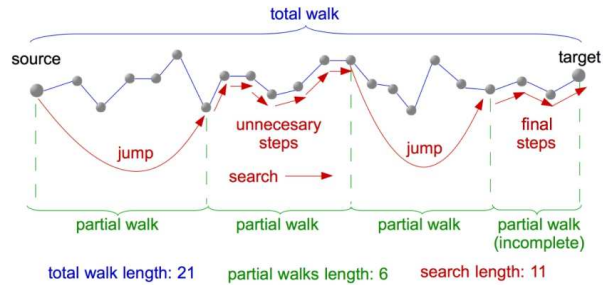}
 \caption{An example search: total walk, partial walks, jumps and steps.}
 \label{fig:concepts}
\end{figure}

At this point, we emphasize the difference between the \emph{search} just defined and the \emph{total walk} that supports it. Indeed, the total walk consist of the concatenation of \emph{partial walks} as defined above. Searches are therefore shorter in length than their corresponding total walks because of the number of steps saved in jumps over PWs in which we know that the resource is not located, although these saving may be reduced by the unnecessary steps due to outdated information within PWs.

More formally, we describe how searches are performed as follows. Let a search start at a node $A$. A PW in $\W_A$ is chosen uniformly at random. Its associated resource information collected in stage 1 is then queried for (any instance of) the desired resource. 

\begin{itemize}
\item
If the query result is negative (i.e., the resource is not in the set of identifiers associated with that PW), the search \emph{jumps} to node $B$, the last node of that PW. Note that the current node and the node to which the search \emph{jumps} are not neighbors in the overlay network in general. Jumps therefore make use of the underlying communication network.
The process is then repeated at $B$ and the search keeps jumping in this way while the results of the queries are negative. 

\item
If the query result is positive (i.e., the resource is in the set of identifiers), the search \emph{traverses} that PW looking for the resource. It starts checking if the current node has the desired resource. If it does not, the search takes a \emph{step} to the next node of the PW, checking again if it has the resource. The search proceeds through the PW in this way until the resource is found or the PW is finished.

\begin{itemize}
\item
If the resource is found, the search stops. 

\item
Otherwise (i.e., the search reaches the last node of the PW without having found the resource in the previous nodes), it means that the information collected in stage 1 for that PW and the resource of interest is no longer valid. The search considers that the result is negative, and the search process is repeated at the last node of the PW.

\end{itemize}

\end{itemize}

\end{enumerate}

\begin{figure}
 \centering
 \includegraphics[width=8cm,angle=270]{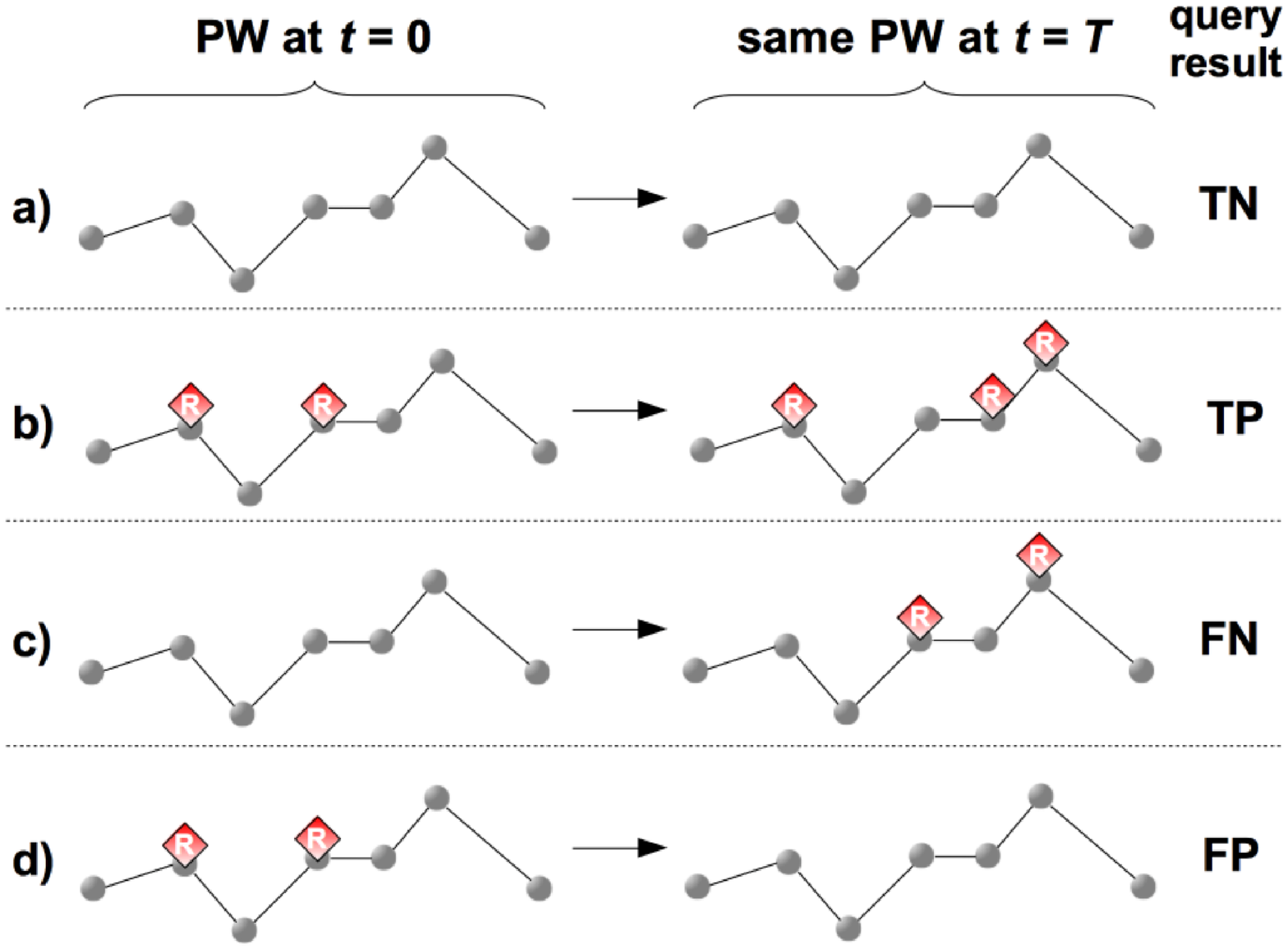}
 \caption{Resource dynamics: examples of results when a PW precomputed at $t\!=\!0$ is queried at $t\!=\!T$.}
 \label{fig:resources_dyn}
\end{figure}

\paragraph{Resource dynamics} Regarding resource dynamics, we realize that searches are executed based on information collected at $t\!=\!0$ that may be outdated during the interval $0 < t \leq T$, when the queries are performed. 
The duration of the interval ($T$) is assumed to be much longer than the duration of a typical search, since we are interested in using the PWs computed at $t=0$ for as many searches as possible, with an acceptable degradation of information associated with those PWs.\footnote{A discussion of the optimal duration of the interval is provided in Section~\ref{sec:cost}.} Therefore, most of the searches take place within the interval, and only a few start in one interval and finish in the following interval. The part of those searches in the following interval benefits from the newly computed PWs of this interval. Our analysis will consider that all searches take place within the interval, thus reflecting a worst case situation.  
Four cases arise when the information associated with a PW is queried for the resource, as is shown in Table~\ref{tab:query_results} and illustrated by Figure~\ref{fig:resources_dyn}. 

A \emph{True Negative} or TN (case a) occurs when no instances were present in the PW at $t\!=\!0$, and the same holds at \mbox{$t\!=\!T$}. A \emph{True Positive} or TP (case b) occurs when one or more instances are present in the PW at $t\!=\!0$ and one or more instances (not necessarily the same ones) are present at $t\!=\!T$. The impact of resource dynamics on the performance of the search mechanism comes from the \emph{False Negatives} (FN) and the \emph{False Positives} (FP). A FN (case c) occurs when there were no instances in the PW at $t\!=\!0$, but at least one instance is present at $t\!=\!T$. It makes the search jump over that PW, ignoring the new instance(s). An FP (case d) occurs when there were one or more instances in that PW at $t\!=\!0$ but all of them are gone at $t\!=\!T$ \emph{and} there are no new instances at $t\!=\!T$. It makes the search traverse a whole PW fruitlessly, since no instances are currently in that PW. Note that the case when all instances disappear from the PW, but some other instance(s) appears in that PW is included in the TP case.

\begin{table}
\centering
\begin{tabular}{ccccc} \hline
     & \multicolumn{2}{c}{\rule{0pt}{11pt}Resource present in PW} &  & \\ \cline{2-3}
case & \rule{0pt}{11pt} $t=0$ & $t=T$ &  \multicolumn{2}{c}{query result}  \\ \hline
a)   & no  & no  & True  Negative & (TN) \\
b)   & yes & yes & True  Positive & (TP) \\
c)   & no  & yes & False Negative & (FN) \\
d)   & yes & no  & False Positive & (FP) \\
\hline
\end{tabular}
\caption{Resource dynamics: query results.}
\label{tab:query_results}
\end{table}

At this point, we note that the performance of the search mechanism can be affected by the degradation of the information collected in the PWs. For instance, if all the instances of a given resource disappear and 
new instances appear in nodes that happen to be in PWs which did not have the resource at $t=0$, the resource would not be found, since searches would jump over those PWs due to the (false) negative result of the queries.
Therefore, in order to keep their accuracy, PWs may need to be ``refreshed''. 
Once the interval is finished, PWs are recomputed at $t\!=\!T$ to acquire fresh resource information, and searches from $t\!=\!T$ to $t\!=\!2T$ will use the new PWs. In other words, stages 1 and 2 described above are repeated with a period $T$. In Section~\ref{sec:cost}, we assess the impact of this PW recomputation on the overall searches cost.

\paragraph{The search metrics}

In this work, we are interested in the number of hops to find a resource, which is defined as the \emph{search length}. This search length is a random variable that takes different values when independent searches are performed. The \emph{search length distribution} is defined as the probability distribution of the search length random variable. So, the \emph{expected search length}, derived from the mentioned distribution, taken over all networks, all source nodes, and target nodes, is an interesting performance measure of the searching mechanism in a given network. 

In the next section we provide an analytical model that gives an estimation of the expected search length of the \emph{choose-first PW-RW} mechanism. We use a \emph{mean-field} model and analysis. In this approach, instead of computing the search length distribution for each network, source node and target node, and then taking expectations, the analysis itself handles average values. Whereas this approach may not accurately reflect some border cases (like some specific network topologies), it captures rather accurately the general characteristics and metrics of the mechanisms. This fact has been also confirmed by using simulations.

\subsection{Analysis}
\label{subsec:analysis}

Since we have defined the expected search length for \emph{any} pair of source and target nodes, 
the expected length of the search from the source node is the same as the expected search length from any subsequent node in the search. This allows us to write a recursive expression for the expected search length. Consider the search is at some node $i$, one of whose PWs is queried for the resource. The expected search length from $i$ can be written in terms of the expected search length from the next node $j$ that will be queried for the resource, \emph{plus} the number of hops taken from node $i$ to node $j$. If the result of the query at $i$ was negative (either false or true), the search will jump from $i$ to $j$, adding 1 hop to the search. If the result was a false positive, the search will traverse the entire PW without finding the resource, adding $s$ hops to the search. Finally, if the result was a true positive, the search will traverse the PW until the resource is found, adding a number of hops whose expected value we call $\Texp$. These cases occur with probabilities that we call, respectively: \pn, \ps, and \pf.
Denoting the expected search length using PWs of length $s$ by \Lsexp, the resulting recursive expression is then:
\begin{equation}
 \Lsexp = (\Lsexp + 1)\cdot \pn + (\Lsexp + s)\cdot \pf + \Texp\cdot \ps,
 \label{eq:Lsexp_SAW_recur}
\end{equation}

\noindent
where \pn, \ps, and \pf\ are the probabilities of choosing a partial walk (out of the \w\ PWs of the node) for which the query for the resource returns a negative (either a TN or a FN, see Table~\ref{tab:query_results}), a TP, and an FP result, respectively, with $\pn+\ps+\pf=1$. \Texp\ is the expectation of the number of final steps taken when traversing the last PW, until an instance of the resource is found (TP).
Solving for \Lsexp, we obtain:
\begin{equation}
 \Lsexp = \frac{1}{\ps} \cdot (\pn + s\cdot\pf) + \Texp.
 \label{eq:Lsexp_RW}
\end{equation}

The probabilities in Equation~\ref{eq:Lsexp_RW} are estimated with the following expressions:

\begin{eqnarray}
  \ps \!\! & = \!\! & \sum_{i=1}^{w} \sum_{j=0}^{w-i} P(i,j)\cdot \frac{i}{w}, \nonumber \\
  \pf \!\! & = \!\! & \sum_{i=0}^{w-1} \sum_{j=1}^{w-i} P(i,j)\cdot \frac{j}{w}, \nonumber \\
  \pn \!\! & = \!\! & \sum_{i=0}^{w-1} \sum_{j=0}^{w-i-1} \!\! P(i,j)\cdot \frac{w-(i+j)}{w} = 1-\ps-\pf,
 \label{eq:RW_probs}
\end{eqnarray}

\noindent where $P(i,j)$ is the probability that, in the \w\ PWs of a node, there are $i$ PWs whose queries return a TP result and $j$ PWs that return an FP result:
\begin{equation}
 P(i,j) = B(\w,\ptp,i)\cdot B(\w-i,\pfp,j),
 \label{eq:Pij}
\end{equation}

\noindent
where 

\begin{itemize}
\item
$B(m,q,n)$ is the coefficient of the binomial distribution
\begin{equation}
B(m,q,n) = \left(\begin{array}{c} m\\ n \end{array}\right)\cdot q^n \cdot (1-q)^{(m-n)},
\end{equation}

\item
\ptp\ is the probability that a given PW at any node returns a TP result,\footnote{This probability \ptp\ is not to be confused with \ps\ (note the different case), defined above as the probability of \emph{choosing} a PW which returns a TP out of the \w\ PWs of the current node.}

\item
and  \pfp\ is the probability that a given PW at any node returns an FP result, conditioned on the fact that it does not return a TP. 

\end{itemize}

Therefore, in order to evaluate the estimation of the expected search length given by Equation~\ref{eq:Lsexp_RW}, we need to obtain the values of \ptp, \pfp\ and \Texp. Let us provide them:

\begin{enumerate}
\item

The variable \ptp\ has been defined as the probability that a given PW at any node returns a TP result. This probability can be easily estimated if we condition it on the fact that the PW (of length \s) had exactly $r$ instances of the resource at $t\!=\!0$. Defining \PWx{r}\ as the probability that a PW has $r$ instances of the resource, and recalling that \R\ is the expectation of the number of instances of the resource in the network, we can write:

\begin{equation}
 \ptp = \!\! \sum_{r=1}^{\mathrm{min}\{\s,\R\}} \PWx{r} \cdot \left[(1-\pd^r) + d^r\cdot\left(1-(1-\pa)^{s-r}\right)\right],
 \label{eq:ptp}
\end{equation}

\noindent where the brackets contain the probability that not all the $r$ instances present at $t\!=\!0$ have disappeared (with probability \pd) at $t\!=\!T$ or, if they did disappear, at least one instance appeared (with probability \pa) in some of the $s-r$ remaining nodes in that interval.

An estimation for \PWx{r}\ can be obtained using the random properties of a random walk in networks built randomly. In particular, we consider that the next hop of a random RW can take it to any of the endpoints in the network (except the endpoints of the current node since we do not allow self-loops). Then we estimate \PWx{r}\ as $B(s,p_{rw},r)$, where $p_{rw}$ is the probability that the RW visits a node with an instance of the resource in the next hop. In turn, we estimate this probability as:

\begin{equation}
 p_{rw} = \frac{\R\cdot\kave}{S-\kRWexp} \cdot \frac{\kRWexp-1}{\kRWexp}.
 \label{eq:prw}
\end{equation}

\noindent The first fraction in Equation~\ref{eq:prw} is the ratio of positive endpoints (the ones connected to the \R\ nodes that have an instance of the resource) and all endpoints in the network ($S=\sum_k k\,\nk$) except those of the current node. We use the average degree of the network ($\kave=\sum_k k\,\nk/N$) as an estimation of the degree of a node that holds the resource (which is assigned or not with uniform probability $\pnoderes$ across the network). Similarly, we use the expectation of the degree of a node visited by a random walk as an estimation of the degree of the current node:

\begin{equation}
 \kRWexp = \sum_k k\cdot \frac{k\cdot n_k}{S} = \frac{1}{S} \cdot \sum_k k^2\cdot n_k.
 \label{eq:pkexp}
\end{equation}

\noindent The second fraction in Equation~\ref{eq:prw} corrects the previous ratio taking into account that, when at a node of a given degree, the probability of not going backwards (and therefore having the chance to find the resource) is the probability of selecting any of its endpoints but the one that connects it with the node just visited by the walk.
With this, the estimation of \PWx{r} is:

\begin{equation}
 \PWx{r} = \left(\begin{array}{c} s\\ r \end{array}\right) \cdot (p_{rw})^r \cdot (1-p_{rw})^{s-r}.
 \label{eq:Prw} 
\end{equation}

\item

We have defined \pfp\ as the probability that a given PW at any node returns an FP result, conditioned on the fact that it does not return a TP. This conditioning comes from the second binomial coefficient in this equation, which we restrict to the $\w-i$ PWs which we know that do not return a TP, since the ones that do are accounted for in the first binomial coefficient. In other words, the second binomial coefficient includes the PWs that return a TN, a FN or an FP result, and \pfp\ is the probability that it returns an FP conditioned on that. We can then easily write an estimation of \pfp\ as:

\begin{equation}
 \pfp = \frac{1}{1-\ptp} \left(1-\PWx{0} - \ptp \right),
 \label{eq:pfp} 
\end{equation}
\noindent where we are substracting \PWx{0}\ (the probability of cases TN and FN in Table~\ref{tab:query_results}), and \ptp\ (the probability of case TP).

\item

An expression for \Texp, the expectation of the number of final steps taken when traversing the last PW until an instance of the resource is found, is still needed to be finally able to estimate the average search length in Equation~\ref{eq:Lsexp_RW}. For this we rely on $\Texpi{r}$, the expectation of that variable conditioned on there being $r$ instances of the resource in the PW. Then:

\begin{equation}
 \Texp = \frac{1}{1-\PWx{0}} \cdot \sum_{r=1}^{\mathrm{min}\{\s,\R\}} \Texpi{r} \cdot \PWx{r}. 
 \label{eq:Texp}
\end{equation}

\noindent Note that \Texp\ is in fact conditioned on there being at least one instance of the resource in the PW, since it corresponds to a TP (see Equation~\ref{eq:Lsexp_SAW_recur}). This is the reason of the fraction multiplying the summation in Equation~\ref{eq:Texp}.

Now we provide an expression for \Texpi{r}\ as the expectation of the position of the first resource in the PW (conditioned on there being $r$ instances of the resource in the PW):

\begin{equation}
 \Texpi{r} = \sum_{i=0}^{s-r} \left[ i \cdot \left( \prod_{j=0}^{i-1} \left( 1-\frac{r}{s-j} \right)  \right) \cdot \left(\frac{r}{s-i}\right) \right].
 \label{eq:Texpi}
\end{equation}

\noindent Each factor in the product of Equation~\ref{eq:Texpi} is the probability that there is no instance in the $j_{th}$ position of the PW, conditioned on that there is no instance in the previous position. The final factor outside the product is the probability that there is an instance in the $i_{th}$ position conditioned on there are no instances in the previous positions. 

\end{enumerate}

\subsection{Performance Evaluation}

In this section, we apply the model presented in the previous section to real networks, and we also validate its predictions with data obtained from simulations. Three types of networks have been chosen for the experiments: regular networks (constant node degree), \ER\ (ER) networks and scale-free networks (with power law on the node degree). These topologies cover a wide variety of real networks~\cite{Albert2002}, ranging from communication networks~\cite{Aiello:2000:RGM:335305.335326} to Internet~\cite{Pastor-Satorras:2004:ESI:1076357} and P2P networks~\cite{Jovanovic:2001}.

A network of each type and size $N=10^4$ has been randomly built with the method proposed by Newman et al.~\cite{nets:Newman01} for networks with arbitrary degree distribution, setting their average node degree to $\kave=10$. Each network is constructed in three steps: (1) a preliminary network is constructed according to its type; (2) its degree distribution is extracted, and (3) the final network is obtained feeding the referenced method with that degree distribution. For each experiment, $10^6$ searches have been performed. In every search, the source node has been chosen uniformly at random, and every node in the network has been assigned an instance of the resource looked for with probability $\pnoderes=10^{-2}$ at $t\!=\!0$. Therefore, the expected number of resource instances present in the network at $t\!=\!0$ for all searches is $\R=\N \cdot \pnoderes = 100$.

We show the results for scale-free networks, which are especially interesting since this type of degree distribution is frequently found in real networks.  Results for regular and ER networks, which do not differ significantly from scale-free results, are included in the Appendix for the interested reader. \\

\subsubsection{Expected search length vs. PW length}

\begin{figure}
 \centering
 \includegraphics[width=8cm]{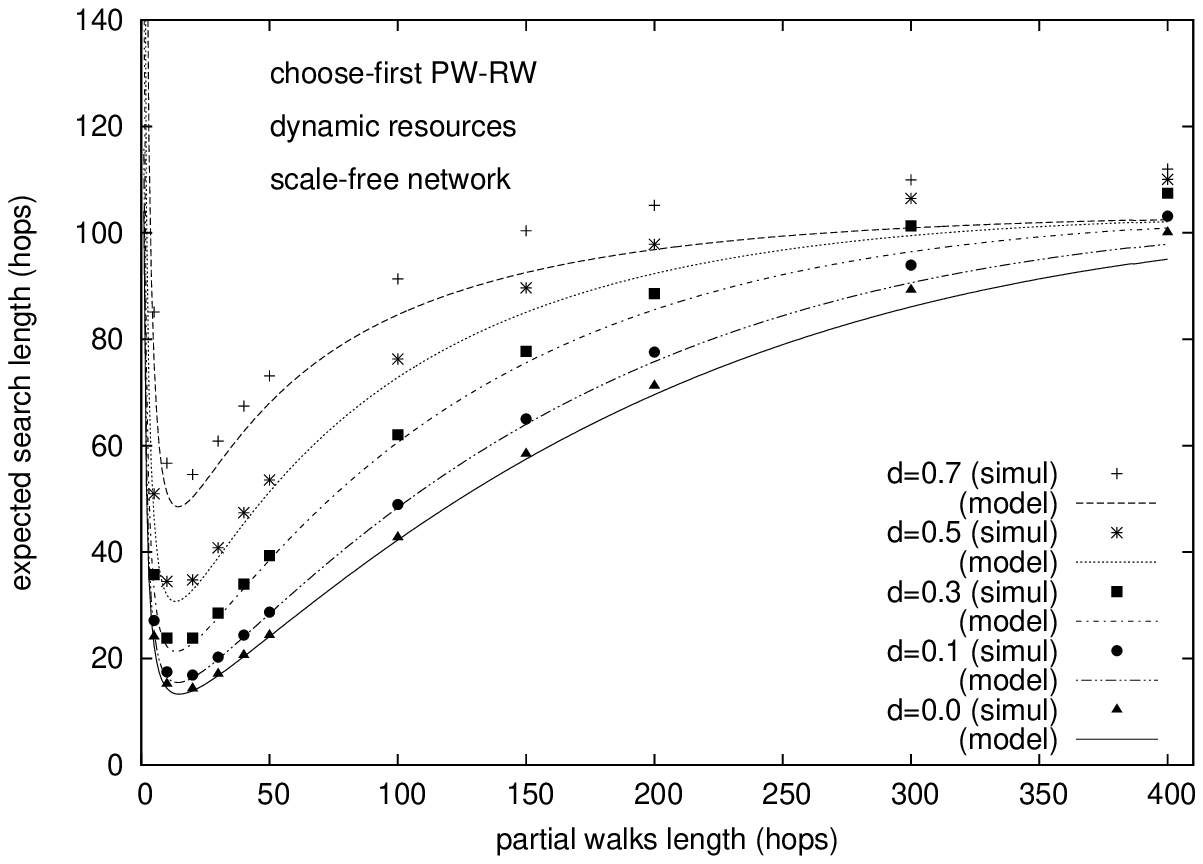}
 \caption{\emph{Choose-first PW-RW} with dynamic resources: expected search length $\Lsexp$ vs. PW length $\s$ for several $\pd$ in a scale-free network.}
 \label{fig:scalefree_pw1}
\end{figure}

Figure~\ref{fig:scalefree_pw1} shows the expected search length in a scale-free network for several values of \pd\ (i.e., the resource dynamics).
The number of PWs per node is set to $\w=5$, although the performance of the PW-RW mechanism is independent from this parameter,\footnote{
The extreme case of having just \emph{one} PW is to be avoided because it yields many unfinished searches, since it is relatively easy to build walks that are loops that do not cover the network. Indeed, if the last node of a PW is a node whose (only) PW has been already used in that total walk, it will take the search to the same place again, resulting in a never-ending loop.} since only one PW is used to locate the resource.
Model predictions (Section~\ref{subsec:analysis}) are plotted with lines and simulation results are shown as points. It can be seen that the model provides an accurate correspondence with the real data, with larger error for higher values of \pd\ and \s. These deviations are discussed further in Section~\ref{sssec:deviations}.

All curves show a minimum point, which marks the optimal PW length (\sopt) and the corresponding optimal expected search length (\Lexpopt). Interestingly, the values of \sopt\ are small and do not depend heavily on \pd. According to the analytic data, \sopt\ ranges between 13 and 15 for all shown curves. \\

\subsubsection{Reduction of the expected search length}

We have compared the performance of the proposed search mechanism for \Lexpopt\ with searches based on simple random walks (RW searches), finding large reductions in the average search lengths. Table~\ref{tab:H_reduction_RW_PWRW} shows the relative reductions (\%) for several values of \pd.  We can see that the reductions that \emph{choose-first PW-RW} achieves with respect to RW searches are lower for higher \pd, ranging from around 88\% in the case when \pd\ = 0 to 57\% when \pd\ = 0.7.

\begin{table}
\centering
\begin{tabular}{|c|c|}
\hline
\rule{0pt}{11pt}   & search length \\ 
              $d$  & reduction (\%)\\ \hline
              0.0  &  88.28 \\
              0.1  &  86.37 \\
              0.3  &  81.23 \\
              0.5  &  72.94 \\
              0.7  &  57.26 \\
\hline
\end{tabular}
\caption{\emph{Choose-first PW-RW} with dynamic resources: reduction of the expected search lengths relative to RW searches for several \pd.}
\label{tab:H_reduction_RW_PWRW}
\end{table}

\subsubsection{Deviations of the predictions of the analytical model}
\label{sssec:deviations}

It can be concluded from Figure~\ref{fig:scalefree_pw1} that the proposed analytical model succeeds in capturing the general behavior of the search mechanism observed in the simulations. There are, however, significant deviations that deserve further evaluation and discussion. We look first at the region of the graphic corresponding to interesting practical scenarios, i.e., values of \s\ close to \sopt. For $\s=20$, relative deviations range between 2.9\% ($\pd=0.0$) and 7.9\% ($\pd=0.7$).
We now look at the region for large \s, to explore the dependency of the deviation on the length of the PWs. For $\s=300$, relative deviations are slightly larger, ranging between 3.6\% and 8.2\%.

These deviations may be explained by one aspect of the construction of PWs that is not captured by our analytical model: revisits within a PW. That is, when constructing a PW, the next node visited by the random walk can be one already in that PW. This will in general happen for several nodes of the PW, with increasing probability for longer PWs (larger \s).
Now consider a search traversing a PW with revisited nodes. In the calculation of the probability that a PW contains the resource, our analysis considers each node of the PW independent from the others, which is clearly not true for a revisited node (which appears more than once in the PW but is in fact the same node). This makes our model slightly \emph{optimistic} because it overestimates the probability of a PW containing the resource, and therefore underestimates the probability of a query to that PW resulting in a false positive when resources disappear. This means that there will actually be more PW traversals caused by false positives than expected, increasing the average search length in simulations. This effect is larger for higher \pd\, when more resources disappear and therefore more than expected PWs cause false positive queries, which is in accordance with Figure~\ref{fig:scalefree_pw1}.

\section{Check-first PW-RW with Dynamic Resources}
\label{sec:check_first}

This section describes a variation of the mechanism presented in Section~\ref{sec:choose_first}. Suppose the search is currently in a node and it needs to pick one of the PWs in that node to decide whether to traverse it or to jump over it. Recall from stage~2 under paragraph ``The search mechanism'' that the original mechanism first \emph{chooses} one of the PWs at random, and then \emph{checks} its associated information for the desired resource, resembling the behavior of a random walk. The proposed variation, on the other hand, reverses the order of these tasks. It first \emph{checks} the associated resource information of \emph{all} the PWs of the node, and then randomly \emph{chooses} among the PWs with a positive result, if any (otherwise, it chooses among all PWs of the node, as the original version). This \emph{check-first PW-RW} mechanism improves the performance of the original (or \emph{choose-first PW-RW}) since the probability of choosing a PW with the resource increases, with no extra storage space cost.

There is another, less important, difference between the algorithms. In the original version, the nodes whose resources were registered in the information associated to the PW ranged between the current node and the one before the last node. In the check-first version, the resource information is registered from the \emph{first} node (the next to the current node) to the \emph{last} node in the PW. This change slightly improves the performance of the new version, since the probability of choosing a PW with the resource increases also in the cases where the resource is held by the last node of the PW. The rest of the operation of the mechanism remains the same. 

\subsection{Analysis}

Most of the analysis provided in Section~\ref{subsec:analysis} is still valid for \emph{check-first PW-RW}. We present here the equations that need to be modified to reflect the new behavior. That is the case of Equations~\ref{eq:RW_probs} for the probabilities of choosing a PW with a TP, FP and negative result, respectively. Their counterparts follow. Remember that $i$ and $j$ represent the number of PWs of the node that return a TP result and an FP result, respectively:

\begin{eqnarray}
  \ps & = & \sum_{i=1}^{w} \sum_{j=0}^{w-i} P(i,j)\cdot \frac{i}{i+j}, \nonumber \\
  \pf & = & \sum_{i=0}^{w-1} \sum_{j=1}^{w-i} P(i,j)\cdot \frac{j}{i+j}, \nonumber \\
  \pn & = & P(0,0) = 1-\ps-\pf, 
 \label{eq:RW_probs_checkfirst}
\end{eqnarray}

The expression for the expectation of the number of final steps taken when traversing the last PW until the resource is found (Equation~\ref{eq:Texp}) is still valid. It uses \Texpi{r}, the expectation of the position of the first resource in the PW, conditioned on there being $r$ instances of the resource in the PW. Its expression (Equation~\ref{eq:Texpi}) needs to be modified, since the range of nodes whose resources are associated with the PW has changed from $[0,s-1]$ to $[1,s]$. The indexes limits and their use in the expression have been updated as necessary in the new expression, which completes the analysis of the \emph{check-first PW-RW} mechanism:

\begin{equation}
 \Texpi{r} = \!\! \sum_{i=1}^{s-r+1} \left[ i \cdot \left( \prod_{j=1}^{i-1} \left( 1-\frac{r}{s-j+1} \right)  \right) \cdot \left(\frac{r}{s-i+1}\right) \right].
 \label{eq:Texpi_checkfirst}
\end{equation}

\subsection{Performance Evaluation}
\label{sec:perf_eval2}

For the performance evaluation of the \emph{check-first PW-RW}, we use the same scenarios used in the performance evaluation of the \emph{choose-first PW-RW}. As earlier, the corresponding figures for regular and ER networks can be found in the Appendix. \\

\subsubsection{Expected search length vs. PW length}

\begin{figure}
 \centering
 \includegraphics[width=8cm]{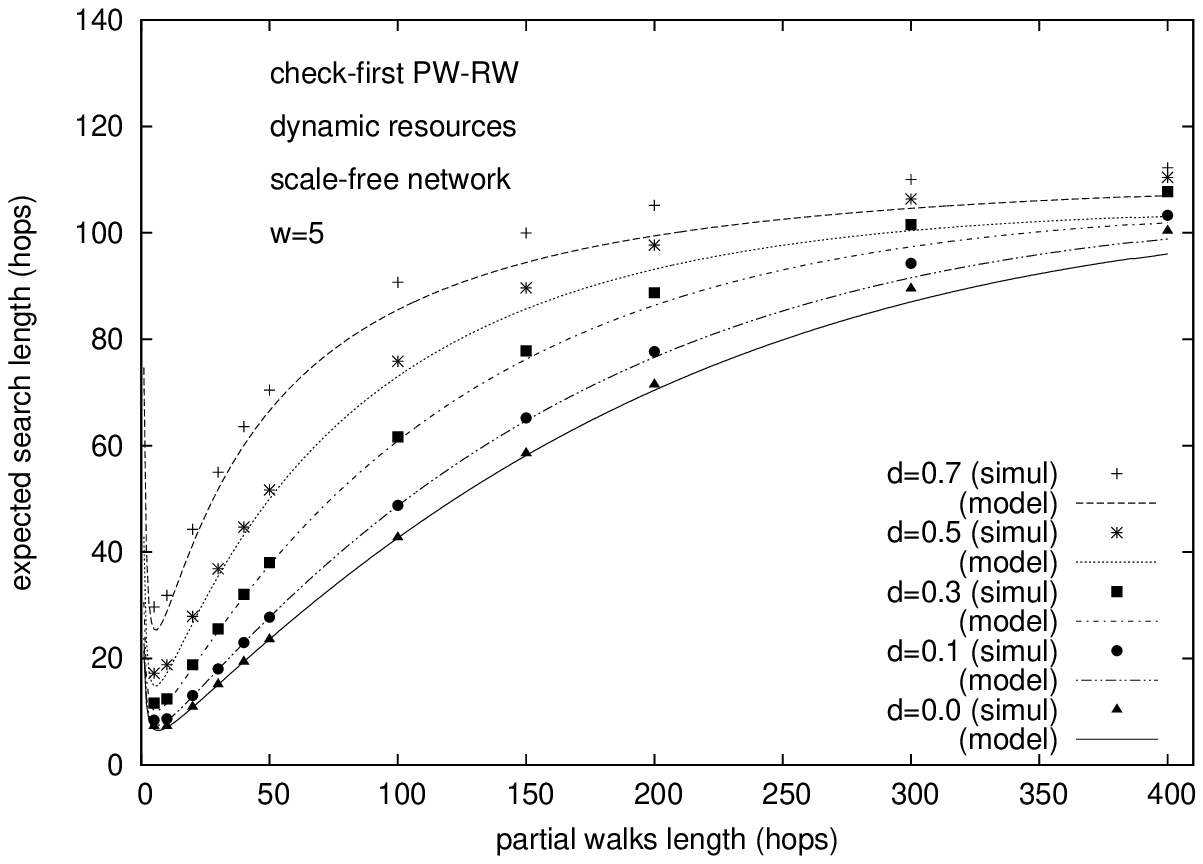}
 \caption{\emph{Check-first PW-RW} ($\w=5$) with dynamic resources: expected search length $\Lsexp$ vs. PW length $\s$ for several $\pd$ in a scale-free network.}
 \label{fig:scalefree_pw2}
\end{figure}

Figure~\ref{fig:scalefree_pw2} shows the expected search length in a scale-free network for several values of \pd, and for $\w=5$. 
The shape of the curves is the same as that for the original mechanism (Figure~\ref{fig:scalefree_pw1}), 
and the discussion on the observed deviations of the analytical results given in Section~\ref{sssec:deviations} is also applicable to this case.

A substantial decrease in the optimal search length (\Lexpopt) for \emph{check-first PW-RW} is observed. For example, for $d=0.3$, \Lexpopt\ is around 11, while it was about 21 for the \emph{choose-first PW-RW} mechanism. The optimal PW length also diminishes, from about 14 to 6 in that case.
The expected search length decrease is due to the fact that the new mechanism checks all the PWs in the node for the resource and then chooses one only among those with positive result, increasing the probability of choosing a PW that currently holds the resource. Following this reasoning, the more PWs in the node, the higher this probability. It is therefore interesting to explore the dependency of \Lexp\ and \sopt\ with \w.

\begin{figure}
 \centering
 \includegraphics[width=8cm]{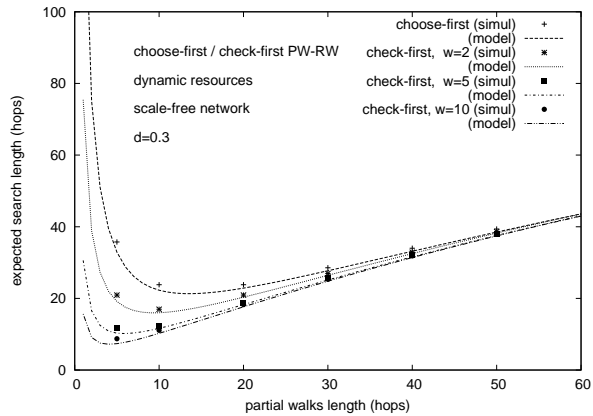}
 \caption{PW-RW with dynamic resources: expected search length $\Lsexp$ vs. PW length $\s$ for \emph{choose-first PW-RW} and \emph{check-first PW-RW} ($\w = 2, 5, 10$) in a scale-free network with $\pd=0.3$.}
 \label{fig:scalefree_pw1_pw2}
\end{figure}

Figure~\ref{fig:scalefree_pw1_pw2} shows the expected search length of the \emph{check-first PW-RW} mechanism, for $\w=2, 5$ and $10$ with $\pd=0.3$. To make the comparison of performance between both mechanisms easier, a curve corresponding to \emph{choose-first PW-RW} has been added to the graph. As explained above, the performance of the latter is independent from \w, but it plays a central role in the \emph{check-first PW-RW}. The range of the axes of this graph has been restricted to focus on the area around \sopt. For higher \s, the curves for the several \w\ converge.
As expected, it is observed that higher \w\ yields lower \Lexpopt, with a value about 8 for $\w=10$. Another interesting observation is that \sopt\ also diminishes for higher \w, falling to 4 in this case. 

These values mean a reduction of about 92\% in the expected search length of simple random walks, with 10 precomputed PWs of just 4 nodes. Higher reductions can be achieved, at the expense of increasing the cost of the computation of the PWs, as analyzed in Section~\ref{sec:cost}. \\

\subsubsection{Reduction of the expected search length}

Table~\ref{tab:H_reduction_RW_PWRW2} is provided as a reference, presenting the reductions achieved by \emph{check-first PW-RW} with respect to random walk searches for $\w=5$  and several \pd. We see that reductions range between 94\% and 78\%, while those of \emph{choose-first PW-RW} ranged between 88\% and 57\%. That is an additional reduction of about 50\%.

\begin{table}
\centering
\begin{tabular}{|c|c|}
\hline
\rule{0pt}{11pt}   & search length \\ 
              $d$  & reduction (\%)\\ \hline
              0.0  &  94.29~\%\\
              0.1  & 93.44~\%\\
              0.3  & 91.04~\%\\
              0.5  & 86.91~\%\\
              0.7  & 78.68~\%\\
\hline
\end{tabular}
\caption{\emph{Check-first PW-RW} ($\w=5$) with dynamic resources: reduction of the expected search lengths relative to RW searches for several \pd.}
\label{tab:H_reduction_RW_PWRW2}
\end{table}

\section{Choose-First and Check-First PW-RW with Dynamic Nodes}
\label{sec:dyn_nodes}

So far we have analyzed the performance of PW-RW when resource instances dynamically appear in, and disappear from, network nodes. In this section, we explore the case where network nodes themselves leave and join the network. In our model, we assume that at time $t\!=\!T$, some nodes have left the network and others have joined it (recall that PWs were computed initially at time $t\!=\!0$).  Nodes that have left the network have a double impact on PW-RW: (1) their resource instances are no longer available, so the information on the resources in a PW is degraded, and (2) searches will not be able to visit them when following their PWs. On the other hand, new nodes have no impact on PW-RW: since they are not in the previously computed PWs, they will not be used by searches, even if they hold instances of the resource looked for.\footnote{New nodes with fresh resource instances will become useful from the searches point of view when PWs are recomputed at $t\!=\!T$.} Therefore, a single parameter $l$, the probability that a node present at $t\!=\!0$ has left at $t\!=\!T$, is enough to evaluate the impact on PW-RW performance at $t\!=\!T$. At this time, the expected number of resource instances is $N\pnoderes(1-l)$, in a network with expected size $N(1-l)$. This implies that the expected number of resource instances per node is still $\pnoderes$, which allows a fair comparison with the baseline (static) case.
The algorithm for the PW-RW mechanisms needs to be adapted for the dynamic nodes scenario. We call nodes that remain in the network at $t\!=\!T$ \emph{active nodes}. When traversing a PW, the walk will visit the first active node at each hop. Similarly, when jumping over a PW, the walk will hop to the last active node in the PW. If no active nodes remain in the PW, the search will proceed visiting any active neighbor or will stop if there are none (only searches that find the resource are included in the performance evaluation results shown later).

\subsection{Analysis}

The analytical models in Sections~\ref{sec:choose_first}~and~\ref{sec:check_first} are modified for the dynamic nodes scenario as follows. 
The expected length of a PW is now $s(1-l)$. Therefore, Equation~\ref{eq:Lsexp_RW} becomes:
\begin{equation}
 \Lsexp = \frac{1}{\ps} \cdot (\pn + s(1-l)\cdot\pf) + \Texp.
 \label{eq:Lsexp_RW_dynnodes}
\end{equation}
Equations~\ref{eq:RW_probs} for \emph{choose-first PW-RW} are still valid, and so are their counterpart Equations~\ref{eq:RW_probs_checkfirst} for \emph{check-first PW-RW}. We need to provide a new version of $\ptp$ in Equation~\ref{eq:Pij}, since the reason of information degradation is now the dynamics of nodes:
\begin{equation}
 \ptp = \sum_{r=1}^{\mathrm{min}\{\s,\R\}} \PWx{r} \cdot (1-l^r),
 \label{eq:ptp_dynnodes}
\end{equation}
with $\PWx{r}$ given by the original Equation~\ref{eq:Prw}, which relied on Equations~\ref{eq:prw}~and~\ref{eq:pkexp}. Equation~\ref{eq:pfp} for $\pfp$ can still be used, with the new definition of $\ptp$ above.
Finally, a new expression for $\Texp$ is needed, since it depends on the number of active nodes $u$ in the last PW, which is now a random variable:
\begin{equation}
 \Texp = \sum_{u=1}^s \Texpi{u} \cdot \PWx{u}, 
 \label{eq:Texp_dynnodes}
\end{equation}
where $\Texpi{u}$ is the expected number of final steps conditioned on there being $u$ active nodes in the PW, and $\PWx{u}$ is the probability that the PW has $u$ active nodes.
To obtain the former we rely on $\Texpi{u,r}$, the expected number of final steps conditioned on there being $r$ instances of the resource in a PW with $u$ active nodes:
\begin{equation}
 \Texpi{u} = \frac{1}{1-\PWx{u,0}} \cdot \sum_{r=1}^{\mathrm{min}\{u,\R\}} \Texpi{u,r} \cdot \PWx{u,r}, 
 \label{eq:Texp_dynnodes2}
\end{equation}
where $\PWx{u,r}$ is the probability of a PW with $u$ active nodes having $r$ instances of the resource. Now, an estimation for $\Texpi{u,r}$ can be obtained reasoning as for Equation~\ref{eq:Texpi}:
\begin{equation}
 \Texpi{u,r} = \sum_{i=0}^{u-r} \left[ i \cdot \left( \prod_{j=0}^{i-1} \left( 1-\frac{r}{u-j} \right)  \right) \cdot \left(\frac{r}{u-i}\right) \right].
 \label{eq:Texpi_dynnodes}
\end{equation}
The expression for $\PWx{u}$ can be obtained as in Equation~\ref{eq:Prw}:
\begin{equation}
 \PWx{u}   = \left(\begin{array}{c} s\\ u \end{array}\right) \cdot (p_{rw,a})^u \cdot (1-p_{rw,a})^{s-u}, \nonumber \\
 \label{eq:Prw2}
\end{equation}
where $p_{rw,a}$, the probability that a node visited by a RW is active, is:
\begin{equation}
 p_{rw,a} = \frac{N(1-l)\cdot\kave}{S-\kRWexp}.
 \label{eq:prwa}
\end{equation}
Likewise, the expression for $\PWx{u,r}$ is:
\begin{equation}
 \PWx{u,r} = \left(\begin{array}{c} u\\ r \end{array}\right) \cdot (p_{rw})^r \cdot (1-p_{rw})^{u-r},
 \label{eq:Prw3}
\end{equation}
where $p_{rw}$ is given by Equation~\ref{eq:prw}.

\subsection{Performance Evaluation}
We look now at the performance of the PW-RW mechanisms in networks with dynamic nodes, that operate according to the model described in the previous section. As earlier, the corresponding figures for regular and ER networks can be found in the Appendix. \\

\subsubsection{Expected search length vs. PW length}

\begin{figure}
 \centering
 \includegraphics[width=8cm]{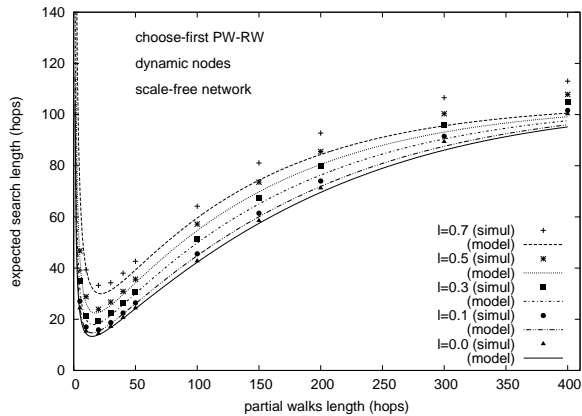}
 \caption{\emph{Choose-first PW-RW} with dynamic nodes: expected search length $\Lsexp$ vs. PW length $\s$ for several $l$ in a scale-free network.}
 \label{fig:scalefree_pw1_dynnodes}
\end{figure}

Figure~\ref{fig:scalefree_pw1_dynnodes} shows results for \emph{choose-first PW-RW} in a scale-free network for several values of $l$, the probability of a node leaving the network. Predictions of the analytical model in Section~\ref{sec:dyn_nodes}, plotted as lines, provide accurate estimations for experimental results, shown as points.
The observed deviations are discussed in Section~\ref{sssec:deviations2}. 
  
The effect of node dynamics is in general very similar to that of resource dynamics (see Figure~\ref{fig:scalefree_pw1}). We note, however, that \sopt\ slightly increases with $l$ in Figure~\ref{fig:scalefree_pw1_dynnodes}, while it slightly decreased for larger $d$ (the probability of a resource instance disappearing) in Figure~\ref{fig:scalefree_pw1}. In addition, we see that \Lexpopt\ remains lower in the dynamic nodes case as $l$ grows, compared to the dynamic resources case for the same values of $d$. Both effects have their origin in the fact that PWs are effectively shorter in the dynamic nodes case, since some of their nodes are no longer in the network when searches are performed. This contributes to shorten the searches, obtaining the optimal lengths with originally longer PWs.

Table~\ref{tab:H_reduction_RW_PWRW_dynnodes} shows the reductions in expected search length achieved by \emph{choose-first PW-RW} relative to RW searches in scale-free networks with dynamic nodes. As for the dynamic resources case (Table~\ref{tab:H_reduction_RW_PWRW}), reductions are similar, ranging in this case between 73\% and 88\%. Reductions for large $l$ are significantly higher than those for large $d$ in the dynamic resources case (about 73\% for $l=0.7$ vs. about 57\% for $d=0.7$). This is explained as above by the effectively shorter PWs in the dynamic nodes case.
\begin{table}
\centering
\begin{tabular}{|c|c|}
\hline
\rule{0pt}{11pt}   & search length \\ 
              $d$  & reduction (\%)\\ \hline
              0.0  &  88.28~\% \\
              0.1  &  87.08~\% \\
              0.3  &  84.21~\% \\
              0.5  &  80.21~\% \\
              0.7  &  73.58~\% \\
\hline
\end{tabular}
\caption{\emph{Choose-first PW-RW} with dynamic nodes: reduction of the expected search lengths relative to RW searches for several $l$.}
\label{tab:H_reduction_RW_PWRW_dynnodes}
\end{table}

\begin{figure}
 \centering
 \includegraphics[width=8cm]{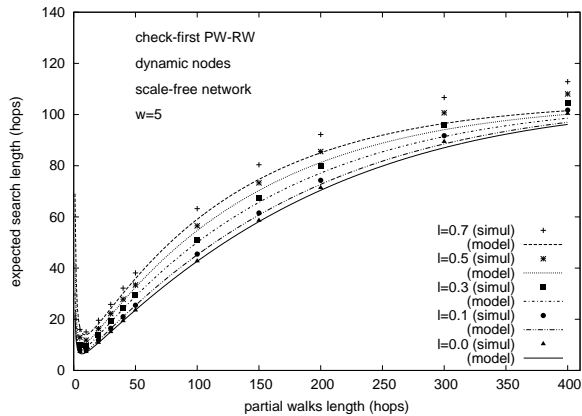}
 \caption{\emph{Check-first PW-RW} ($\w=5$) with dynamic nodes: expected search length $\Lsexp$ vs. PW length $\s$ for several $l$ in a scale-free network.}
 \label{fig:scalefree_pw2_dynnodes}
\end{figure}

\begin{figure}
 \centering
 \includegraphics[width=8cm]{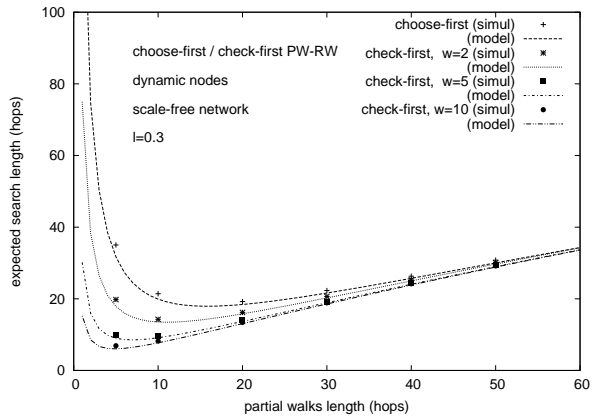}
 \caption{PW-RW with dynamic nodes: expected search length $\Lsexp$ vs. PW length $\s$ for \emph{choose-first PW-RW} and \emph{check-first PW-RW} ($\w = 2, 5, 10$) in a scale-free network with $\pd=0.3$.}
 \label{fig:scalefree_pw1_pw2_dynnodes}
\end{figure}

Figure~\ref{fig:scalefree_pw2_dynnodes} shows the results for the \emph{check-first PW-RW} mechanism in a scale-free network with dynamic nodes. As in the dynamic resources case (see Figure~\ref{fig:scalefree_pw2}), both \sopt\ and \Lexpopt\ decrease with respect to the choose-first mechanism, since the algorithm checks all PWs available at the node for the resource, increasing the probability of finding it. Figure~\ref{fig:scalefree_pw1_pw2_dynnodes} shows that both \Lexpopt\ and \sopt\ decrease for larger $w$ (the number of PWs per node), as observed also for the dynamic resources case (see Figure~\ref{fig:scalefree_pw1_pw2}). \\

\subsubsection{Reduction of the expected search length}

Finally, Table~\ref{tab:H_reduction_RW_PWRW2_dynnodes} shows the reductions in expected search length of \emph{check-first PW-RW} compared to RW searches. Reduction values range between 87\% and 94\%. Again, reductions for high $l$ are larger than in the dynamic resources case for large $d$ (around 87\% for $l=0.7$ vs. 77\% for $d=0.7$). Reductions relative to \emph{choose-first PW-RW} are above 51\%, slightly larger than for the dynamic resource case.
\begin{table}
\centering
\begin{tabular}{|c|c|}
\hline
\rule{0pt}{11pt}   & search length \\ 
              $d$  & reduction (\%)\\ \hline
              0.0  &  94.29~\% \\
              0.1  & 93.76~\% \\
              0.3  & 92.48~\% \\
              0.5  & 90.68~\% \\
              0.7  & 87.73~\% \\
\hline
\end{tabular}
\caption{\emph{Check-first PW-RW} ($\w=5$) with dynamic nodes: reduction of the expected search lengths relative to RW searches for several $l$.}
\label{tab:H_reduction_RW_PWRW2_dynnodes}
\end{table}

\subsubsection{Deviations of the predictions of the analytical model}
\label{sssec:deviations2}

Figures~\ref{fig:scalefree_pw1_dynnodes}~and~\ref{fig:scalefree_pw2_dynnodes} show deviations of the proposed analytical model with respect to simulation results. These deviations are very similar to those for the dynamic resources scenario, discussed in Section~\ref{sssec:deviations}. For example, deviations of the analytical model for \emph{choose-first PW-RW} with dynamic nodes range from 3.2\% ($l=0.0$) to 9.4\% ($l=0.7$) when $\s=20$, and from 3.7\% to 10.3\% when $\s=300$. 
As for the dynamic resources case, these deviations may be caused by revisits within PWs. Our analytical models do not take this effect into account, resulting in optimistic predictions (see Section~\ref{sssec:deviations}). 

\section{Robustness of PW-RW to parameter perturbations}

In this section, we assess the robustness of the proposed mechanisms to variations of some of the parameters, relative both to the network and to the mechanisms themselves. For this, we use the analytical models defined and validated against simulation results in the previous sections. The PW-RW mechanisms have two parameters: the length of the PWs (\s) and the number of PWs per node (\w). The impact of variations of \s\ on the expected search length has already been shown in Figures~\ref{fig:scalefree_pw1}~to~\ref{fig:scalefree_pw1_pw2_dynnodes}, revealing the existence of an optimal length for PWs (\sopt), which achieves minimum expected search length. As for \w, it is only relevant for  \emph{check-first PW-RW}, since all the PWs available in the node are checked for the resource. Figures~\ref{fig:scalefree_pw1_pw2}~and~\ref{fig:scalefree_pw1_pw2_dynnodes} show that the minimum expected search length (and also \sopt) decreases as \w\ is increased, as expected. 

Regarding network parameters, we have already studied the impact of the dynamics of resources (through $d$, the probability that a resource disappears from a node) and of the dynamics of nodes (through $l$, the probability that a node leaves the network), observing that the minimum expected search lengths remain reasonably low even for high volatility probabilities (see Figures~\ref{fig:scalefree_pw1}~to~\ref{fig:scalefree_pw1_pw2_dynnodes}).

\begin{figure}
 \centering
 \includegraphics[width=8cm]{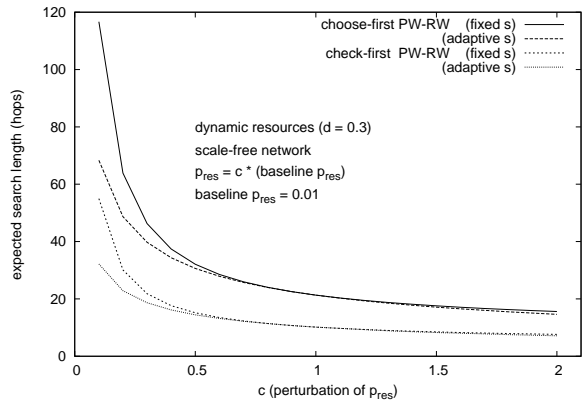}
 \caption{PW-RW with dynamic resources: expected search length $\Lsexp$ vs. $c$, the perturbation of \pnoderes\ in a scale-free network with $d=0.3$.}
 \label{fig:scalefree_perturbation_pres_dynres}
\end{figure}

We now investigate the effect of \pnoderes, the probability that a node has an instance of the resource looked for, which in turn determines the average number of instances in the network. We write \pnoderes\ as a function of a multiplicative perturbation factor $c$ for the chosen baseline value (0.01), i.e., $\pnoderes=0.01\cdot c$. Then we vary $c$ in the range [0.1, 2] and observe the performance of the mechamisms in a scale-free network with dynamic resources ($d=0.3$). We have found that the value of \sopt\ itself varies when \pnoderes\ is perturbed, ranging from 44 to 10 for \emph{choose-first} and from 19 to 4 for \emph{check-first}. We are presented with two options here: calculate the expected search lengths with $s$ fixed to the \sopt\ corresponding to the baseline \pnoderes\ (14 for \emph{choose-first} and 6 for \emph{check-first}), or use the actual \sopt\ for each value of \pnoderes\ (in a hypothetical \emph{adaptive} PW-RW). Figure~\ref{fig:scalefree_perturbation_pres_dynres} compares both cases for \emph{choose-first PW-RW} (upper curves) and \emph{check-first PW-RW} (lower curves). Expected search lengths grow for $c<1$ (as the number of resources decreases below the baseline), and diminish for $c>1$ (as the number of resources increases over the baseline), as expected. More interesting is the fact that the proposed mechanisms (with fixed $s$) and the hypothetical adaptive versions achieve very similar expected search lengths in a wide range of $c$. This suggests that the PW-RW mechanisms are robust to variations of \pnoderes, with little performance degradation due to the suboptimality of $s$ when \pnoderes\ varies around a baseline value. Results are similar for networks with dynamic nodes, and are not shown here.

Finally, we have checked that the expected search lengths (and also \sopt) do not depend on the network size ($N$), as long as \pnoderes\ and the average degree of the network remain unchanged. This is as expected, since fixing \pnoderes\ means that the number of resource instances is proportional to the network size, and fixing the average degree means that the number of choices of a random walk to visit any endpoint in the network (the expected number of neighbors of the current node) remains the same.

\section{Cost of the PW-RW mechanisms}
\label{sec:cost}

In our proposed searching mechanisms, we assumed that at some initial time all network nodes compute a number $w$ of PWs that are used for an interval of time of length $T$. After that, the searches performed in this interval use these PWs, and thus the cost of their computation must be added to the cost of the searches themselves. In this section we show that such an increase is not relevant in the average cost per search, regardless of the value of $T$.

First, we observe that since the information associated with those initial PWs is degraded due to the network dynamics, new PWs are computed at $t\!=\!T$ to capture the current network state. These new PWs are then used for another interval of length $T$, and so on. The election of the length of the PW recomputation interval $T$ influences the total cost with two opposite effects. For longer intervals, more searches share the cost of the computation of PWs, lowering the cost per search. However, the information associated with the PWs gets more degraded for longer intervals, causing longer searches. This suggests the existence of some optimal $T_\mathrm{opt}$ that minimizes the total cost per search.

To investigate this we use $C_t$, the \emph{average total cost per search} for interval length $T$, as defined in~\cite{Lopez:Computing} for the static network case, as the goodness metric to optimize. This total cost, defined as the number of messages, takes into account the cost of the average search (\Lsexp\ messages plus one to notify the source) and the cost of the computation of each of the $w$ PWs of a node ($s$ messages plus one to notify the node), divided by the number of searches $b$ performed per node in the interval:
 
\begin{equation}
 C_t = (\Lsexp+1) + \frac{w}{b}(\s+1).
 \label{eq:Ct}
\end{equation} 

In dynamic networks, however, \Lsexp\ varies with time during the interval as information progresively degrades, so it needs to be averaged over $T$. In addition, since we intend to study the average total cost as a function of $T$, the number of searches per node in the interval depends on $T$ as $b = \lambda T$, if we assume that the number of searches per node per time unit is a constant that we denote $\lambda$. The previous equation then becomes:
\begin{equation}
 C_t(T) = \frac{1}{T}\int_0^T(\Lsexp(t)+1) \cdot dt + \frac{w}{\lambda T}(\s+1).
 \label{eq:Ct_t}
\end{equation}

$\Lsexp(t)$ is determined by the stochastic process of deaths that is governed by $\mu_k$, the rate of departures of resource instances or nodes when the system is in state $k$ (the number of instances or nodes). A rigorous approach to this problem would require a complete stochastic characterization of the behavior of a system with a large number of states, and is out of the scope of this work. For our purpose, we will focus on intervals of length $T$ in which a relatively small fraction of resources or nodes leave the system, so that the expected search length does not overly increase. For these intervals, we will assume that $\mu$ remains constant. This is cleary just an approximation to the behavior of a real system, but it greatly simplifies the analysis allowing us to draw interesting practical conclusions.

To obtain an estimation of $\Lsexp(t)$ under these assumptions, we first seek $\Lsexp(d)$, the relation between the expected search length and the fraction $d$ of resource instances that disappear. Similarly, we need to find $\Lsexp(l)$ for networks with dynamic nodes. The following discussion focuses on the dynamic resources case for conciseness. To find $\Lsexp(d)$, which we note is independent from time, we use the analysis of the PW-RW mechanisms presented in the previous sections.

\begin{figure}
 \centering
 \includegraphics[width=8cm]{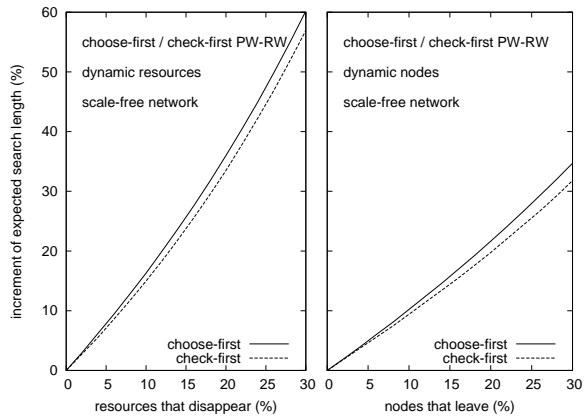}
 \caption{\emph{Choose-first} and \emph{check-first PW-RW} in dynamic networks: relative increment of the expected search length \Lsexp\ vs. (left) the fraction $d$ of resource instances $d$ that disappear, and (right) the fraction $l$ of nodes that leave the network, in a scale-free network.}
 \label{fig:sca_incL_d_l}
\end{figure}

Figure~\ref{fig:sca_incL_d_l} shows the increments in the expected search length for the \emph{choose-first} and \emph{check-first} PW-RW mechanisms in a scale-free network with dynamic resources (left) and dynamic nodes (right). Increments are obtained relative to the expected search length for the static case (e.g., $(\Lsexp(d)-\Lsexp(0))/\Lsexp(0)$). We observe that these dependencies are approximately linear, with a slope $\alpha \approx 2$ for dynamic resources and $\alpha \approx 1$ for dynamic nodes. We can then write $\Lsexp(t) \approx \Lsexp(0) (1+\alpha d(t))$. Now, if $\mu$ is the departure rate of resource instances (or nodes), and assuming it constant for the intervals of interest, we obtain an estimation for the fraction of instances that have disappeared at time $t$ as $d(t)\approx \mu t/N$. With all the above we can rewrite Equation~\ref{eq:Ct_t} as:
\begin{eqnarray}
 C_t(T) \!\! & = \!\! & \frac{1}{T}\int_0^T\left(\Lsexp(0)\cdot(1+\alpha \frac{\mu}{N}t)+1\right) \cdot dt + \frac{w}{\lambda T}(\s+1) \nonumber \\
        \!\! & = \!\! & \left(\Lsexp(0) + 1\right) + \frac{\Lsexp(0)\cdot\alpha\mu}{2N}\ T + \frac{w(s+1)}{\lambda}\ \frac{1}{T}.
 \label{eq:Ct_t2}
\end{eqnarray}

A simple calculation on the previous equation allows us to obtain an expression for the interval $T_{\mathrm{opt}}$ that minimizes the average total cost per search:
\begin{equation}
 T_{\mathrm{opt}} = \sqrt{\frac{2Nw(s+1)}{\Lsexp(0)\cdot \alpha\mu\lambda}}.
 \label{eq:Topt}
\end{equation}

\begin{figure}
 \centering
 \includegraphics[width=8cm]{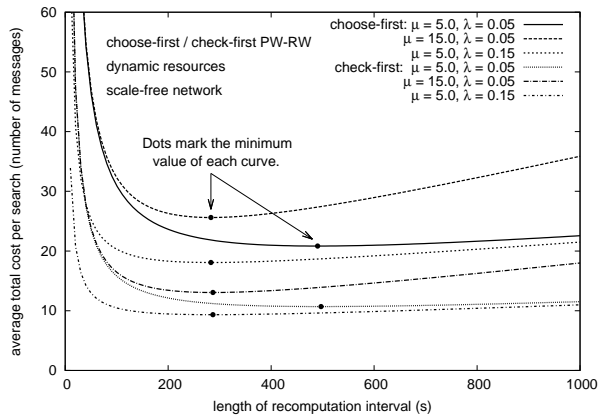}
 \caption{\emph{Choose-first} and \emph{check-first PW-RW} with dynamic resources: average total cost per search $C_t$ vs. the length of the PW recomputation interval $T$, in a scale-free network.}
 \label{fig:sca_dynres_cost_T}
\end{figure}

\begin{figure}
 \centering
 \includegraphics[width=8cm]{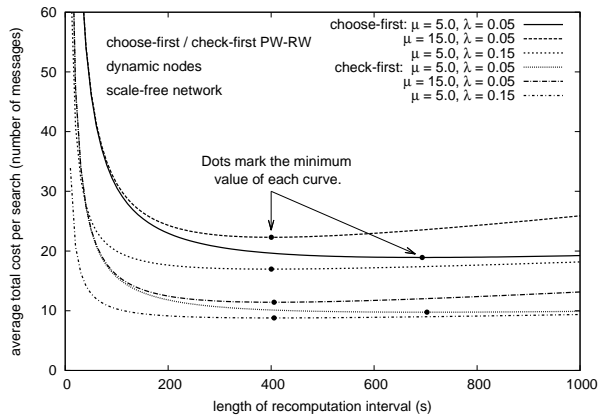}
 \caption{\emph{Choose-first} and \emph{check-first PW-RW} with dynamic nodes: average total cost per search $C_t$ vs. the length of the PW recomputation interval $T$, in a scale-free network.}
 \label{fig:sca_dynnod_cost_T}
\end{figure}

Figures~\ref{fig:sca_dynres_cost_T}~and~\ref{fig:sca_dynnod_cost_T} show plots of the average total cost per search $C_t(T)$ for the \emph{choose-first} and \emph{check-first} PW-RW mechanisms in networks with dynamic resources and with dynamic nodes, respectively, for several values of $\mu$ and $\lambda$. All curves have a linear behavior for large $T$ (where cost is dominated by longer searches), and a rapid increase for small $T$ (where cost is dominated by PW computation). The minimum cost, marked with a dot, can be found in the region in between. Interestingly, this intermediate region is observed to be quite flat, which allows to reduce the interval $T$ below $T_\mathrm{opt}$ with no relevant increase in the average cost per search, but with significant reductions in the expected search length (see Figure~\ref{fig:sca_incL_d_l}). This also means that the PW-RW mechanisms are not very sensitive to the election of the length $T$ of the recomputation interval, given that the system remains in the intermediate region. Finally, the plots of $C_t(T)$ also show that the cost increases for more volatile networks (higher $\mu$) and decreases for more frequent searches (higher $\lambda$), as it was expected.

\section{Conclusions}
\label{sec:conclusions}

We have proposed a mechanism to locate a desired resource in randomly built networks with dynamic resource behavior (resource instances can appear and disappear) or dynamic node behavior (nodes can join and leave the network). The mechanism is based on building a total walk with partial walks that are precomputed as random walks and available at each network node. When precomputing each partial walk, information on the resources held by its nodes is stored and associated to it. This information is used by the searches, so that they can jump over partial walks in which the desired resource is not located. Two versions of the mechanism have been described. In the \emph{choose-first} version, one of the partial walks at the current node is randomly chosen, and then checked for the desired resource. In the \emph{check-first} version, all the partial walks of the node are checked for the resource, and then one is randomly chosen among those in which the resource was found. 
We have presented an analytical model that predicts the expected search length achieved by the two versions of the mechanism. Simulation experiments have been used to validate the model and to assess the effect of resource and node dynamics. We have found that the \emph{choose-first} version achieves large reductions of the average search length in relation to searches based on simple random walks. These reductions remain significant even in the face of high volatility of resources or nodes. The \emph{check-first} version produces larger reductions as we increase the number of partial walks precomputed at each node (at the corresponding extra cost). Results have been found to be very similar for networks with different degree distributions ($k$-regular, \ER\ and scale-free).
Finally, we have analyzed the cost of the PW-RW mechanisms, concluding that the choice of the length of the PW precomputation interval does not have a significant impact on the cost in a wide range of interval lengths. 

An interesting future work line for this study is to measure the improvement in the search length that can be obtained by using different strategies to choose one of the partial walks available in a node. Another possibility to shorten search lengths is to use more intelligent (and more costly) variations of random walks instead of simple random walks.

\newpage

\bibliographystyle{abbrv}
\bibliography{randomwalks,networks}

\newpage

\appendix

This Appendix contains results for regular and ER networks that complement those given for scale-free networks in the main text, as well as some additional results. Results for the three network types are very similar in all cases.

\section{Choose-First PW-RW with Dynamic Resources}

\begin{figure}[h]
 \centering
\includegraphics[width=8cm]{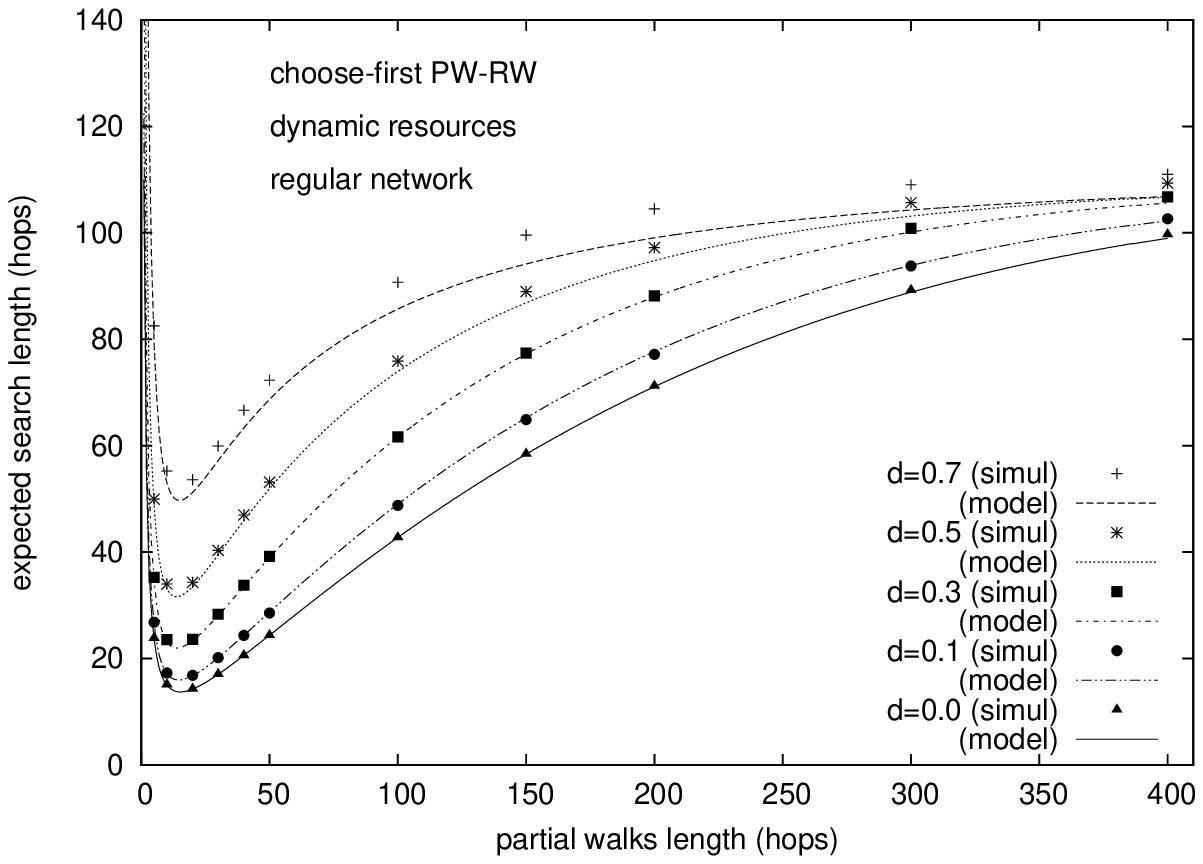}
 \caption{\emph{Choose-first PW-RW} with dynamic resources: expected search length $\Lsexp$ vs. PW length $\s$ for several $\pd$ in a regular network.}
 \label{fig:regular_pw1}
\end{figure}

\begin{figure}[h]
 \centering
 \includegraphics[width=8cm]{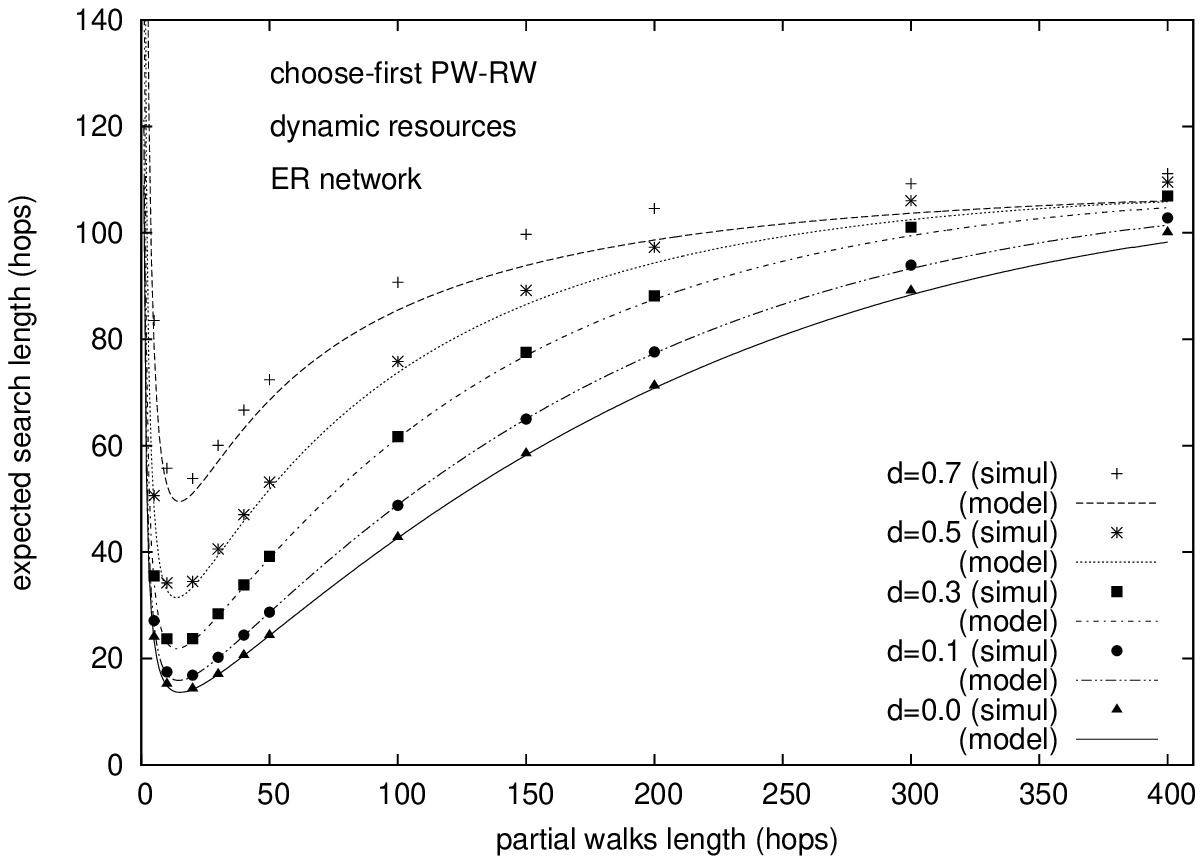}
 \caption{\emph{Choose-first PW-RW} with dynamic resources: expected search length $\Lsexp$ vs. PW length $\s$ for several $\pd$ in a ER network.}
 \label{fig:random_pw1}
\end{figure}

Figures~\ref{fig:regular_pw1}~and~\ref{fig:random_pw1} show experimental and analytical results for the expected search length achieved by \emph{choose-first PW-RW} in regular and ER networks with dynamic resources, respectively, for several values of $p$, the probability of a resource disappearing from a node.

\begin{itemize}
\item{\it Length Distributions and Dispersion:}

The use of partial walks also affects the shape of the probabilistic distribution of search lengths. Figures~\ref{fig:regular_dist}~and~\ref{fig:random_dist} show the distributions for simple random walks (RW) searches and for PW-RW searches, for $\s=10$ and for several values of \pd, obtained from the experiments with regular and ER networks with dynamic resources. Instead of the slowly decaying distribution of RWs, the proposed mechanism exhibits search length distributions that show a maximum frequency for a small search length and then decay much faster than the random walk distribution. We also note that the search length for the maximum frequency (9 in this case) is independent from the dynamic behaviour of resources (\pd). The search length distributions of PW-RW have therefore lower standard deviation than random walk searches. For $d=0.3$, for instance, the standard deviation in regular, ER and scale-free networks are, respectively, 21.06, 21.34 and 21.58. Correspondingly, the mean search lengths are 23.56, 23.72 and 23.81. As a reference for comparison, random walk searches on the scale-free network registered a standard deviation and mean of search lengths of 116.48 and 113.56.

\begin{figure}[h]
 \centering
 \includegraphics[width=8cm]{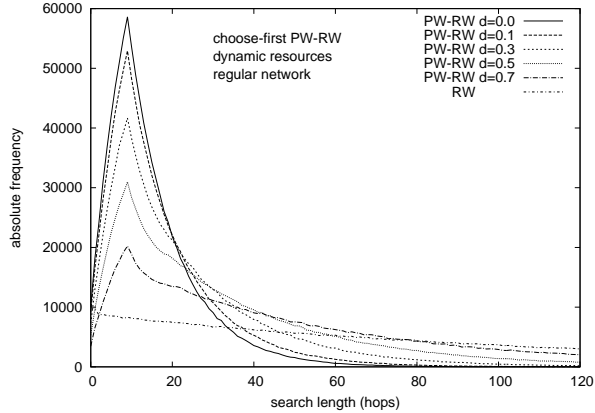}
 \caption{\emph{Choose-first PW-RW} with dynamic resources: search length distributions for RW searches and for PW-RW searches, in a regular network with $\s=10$ and several $\pd$.}
 \label{fig:regular_dist}
\end{figure}

\begin{figure}[h]
 \centering
 \includegraphics[width=8cm]{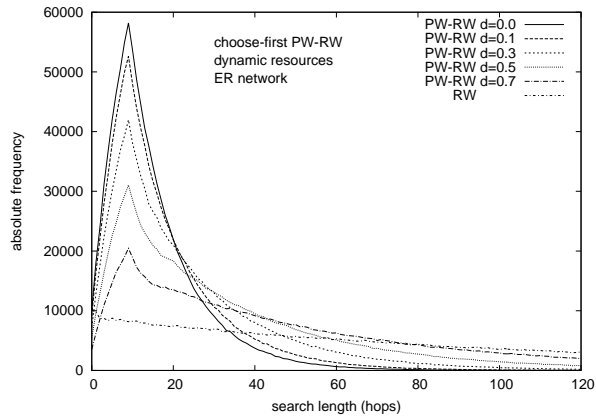}
 \caption{\emph{Choose-first PW-RW} with dynamic resources: search length distributions for RW searches and for PW-RW searches, in a ER network with $\s=10$ and several $\pd$.}
 \label{fig:random_dist}
\end{figure}

\end{itemize}

\section{Check-First PW-RW with Dynamic Resources}

\begin{figure}[h]
 \centering
 \includegraphics[width=8cm]{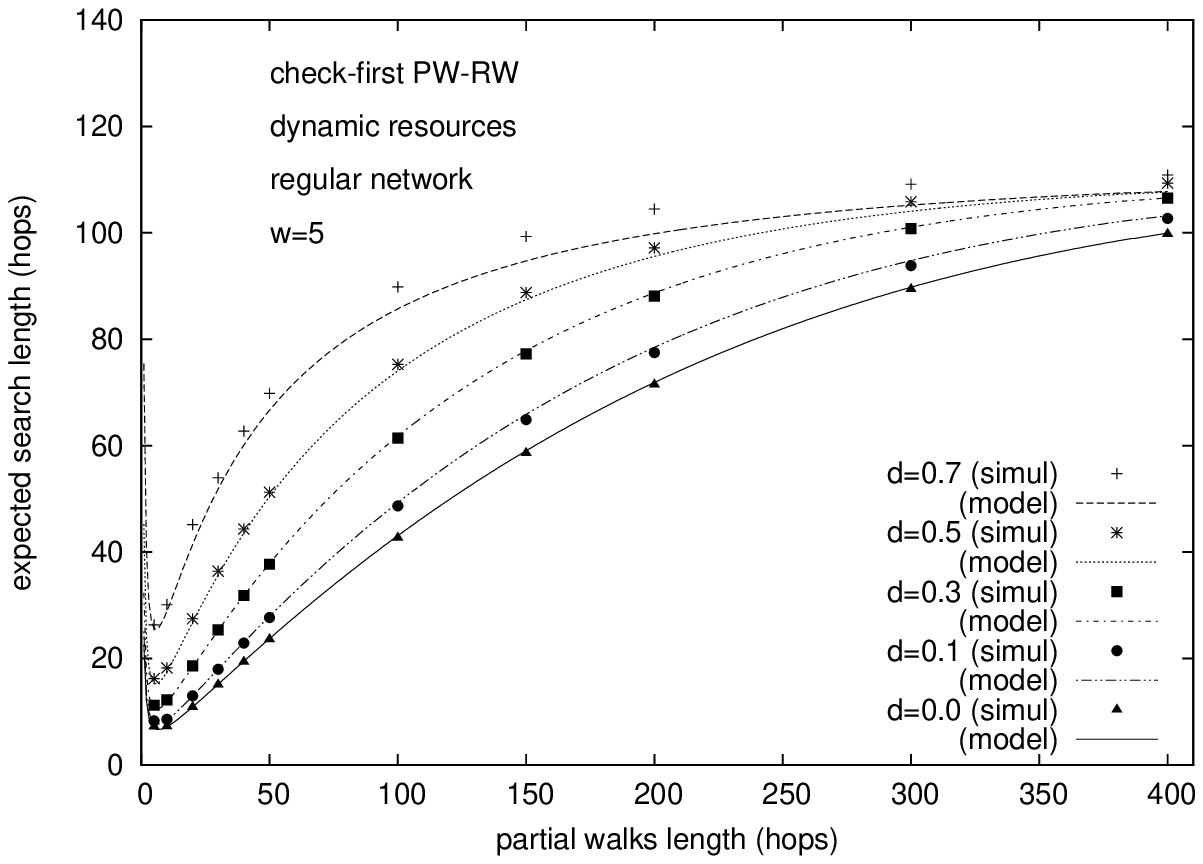}
 \caption{\emph{Check-first PW-RW} ($\w=5$) with dynamic resources: expected search length $\Lsexp$ vs. PW length $\s$ for several $\pd$ in a regular network.}
 \label{fig:regular_pw2}
\end{figure}

\begin{figure}[h]
 \centering
 \includegraphics[width=8cm]{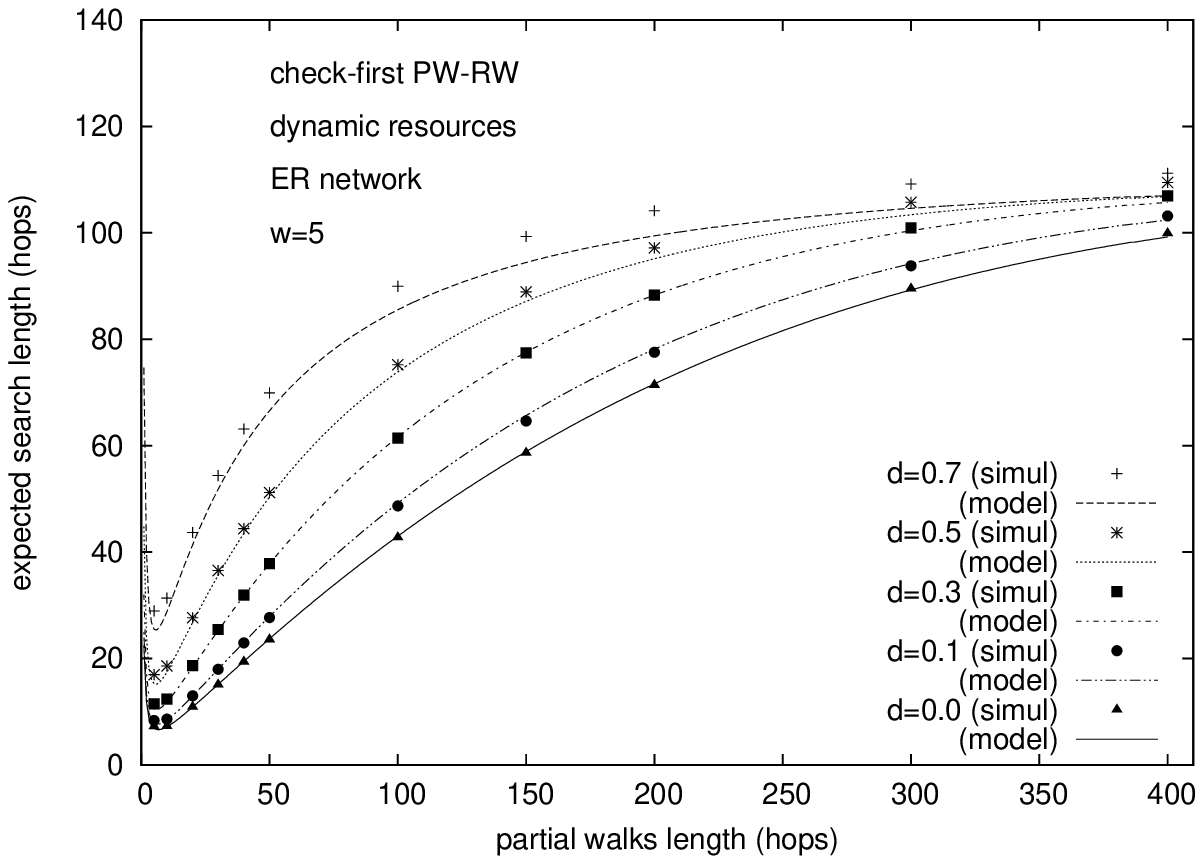}
 \caption{\emph{Check-first PW-RW} ($\w=5$) with dynamic resources: expected search length $\Lsexp$ vs. PW length $\s$ for several $\pd$ in a ER network.}
 \label{fig:random_pw2}
\end{figure}

Figures~\ref{fig:regular_pw2}~and~\ref{fig:random_pw2} show the expected search lengths achieved by \emph{check-first PW-RW} in regular and ER networks with dynamic resources, respectively, for several values of $p$.

\begin{figure}[h]
 \centering
 \includegraphics[width=8cm]{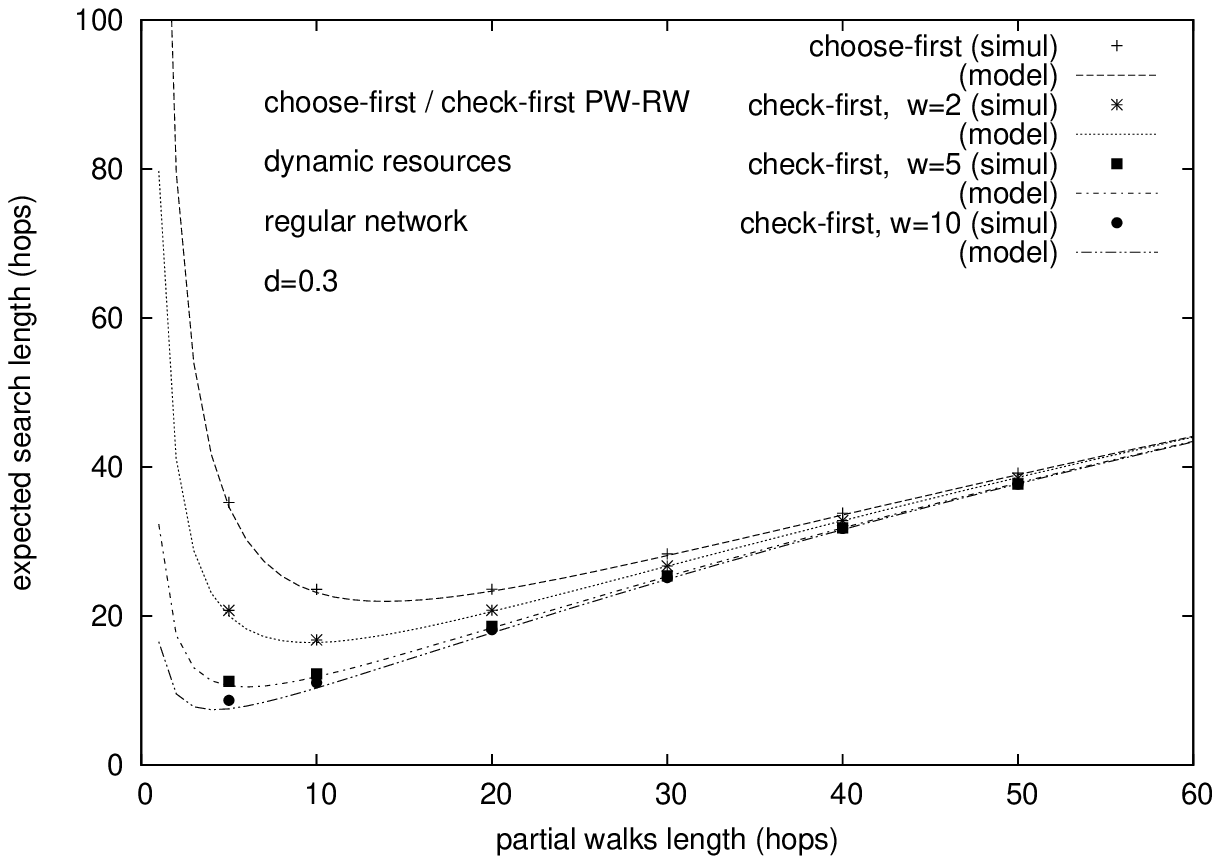}
 \caption{PW-RW with dynamic resources: expected search length $\Lsexp$ vs. PW length $\s$ for \emph{choose-first PW-RW} and \emph{check-first PW-RW} ($\w = 2, 5, 10$) in a regular network with $\pd=0.3$.}
 \label{fig:regular_pw1_pw2}
\end{figure}

In addition, Figure~\ref{fig:regular_pw1_pw2} shows the dependency between the expected search lengths achieved by \emph{check-first PW-RW} and the number of PWs per node ($w$), in a regular network with dynamic resources. Results for ER networks are very similar.

\section{Choose-First PW-RW with Dynamic Nodes}

\begin{figure}[h]
 \centering
 \includegraphics[width=8cm]{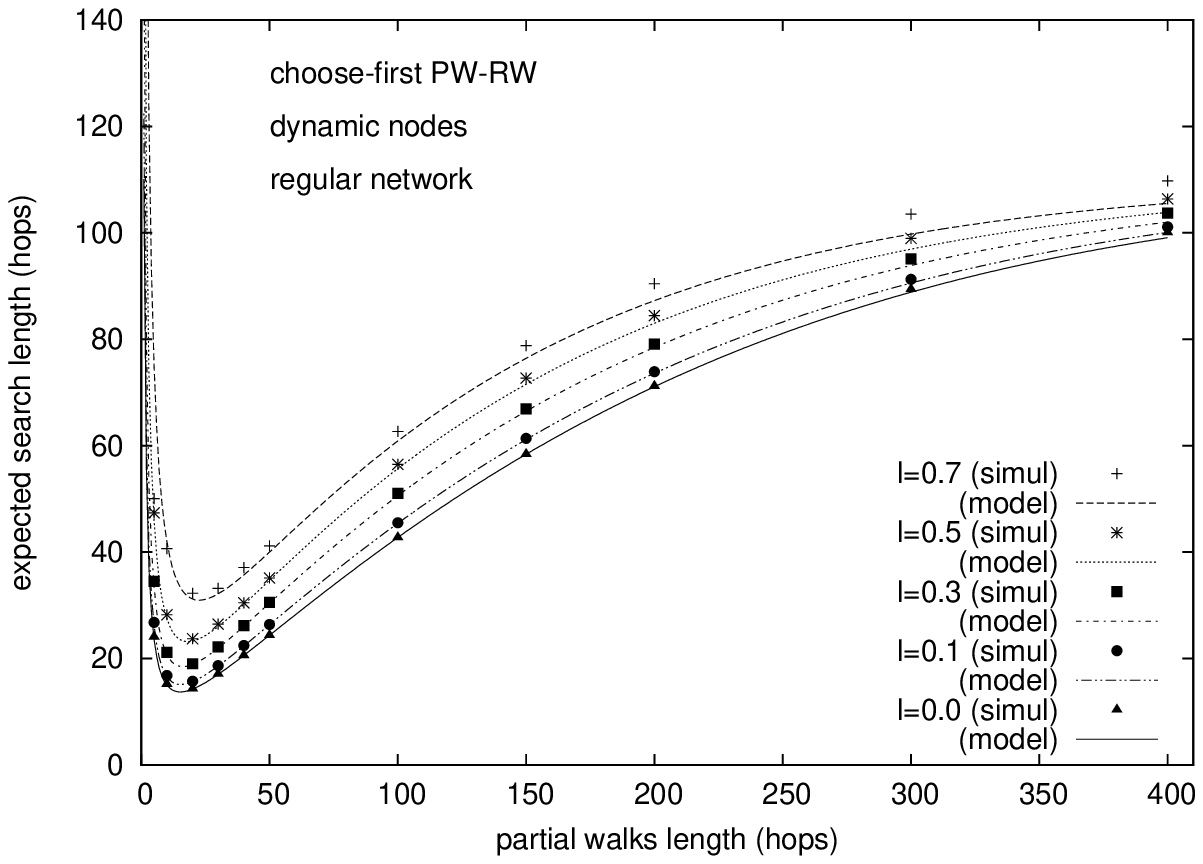}
 \caption{\emph{Choose-first PW-RW} with dynamic nodes: expected search length $\Lsexp$ vs. PW length $\s$ for several $\pd$ in a regular network.}
 \label{fig:regular_pw1_dynnodes}
\end{figure}

\begin{figure}[h]
 \centering
 \includegraphics[width=8cm]{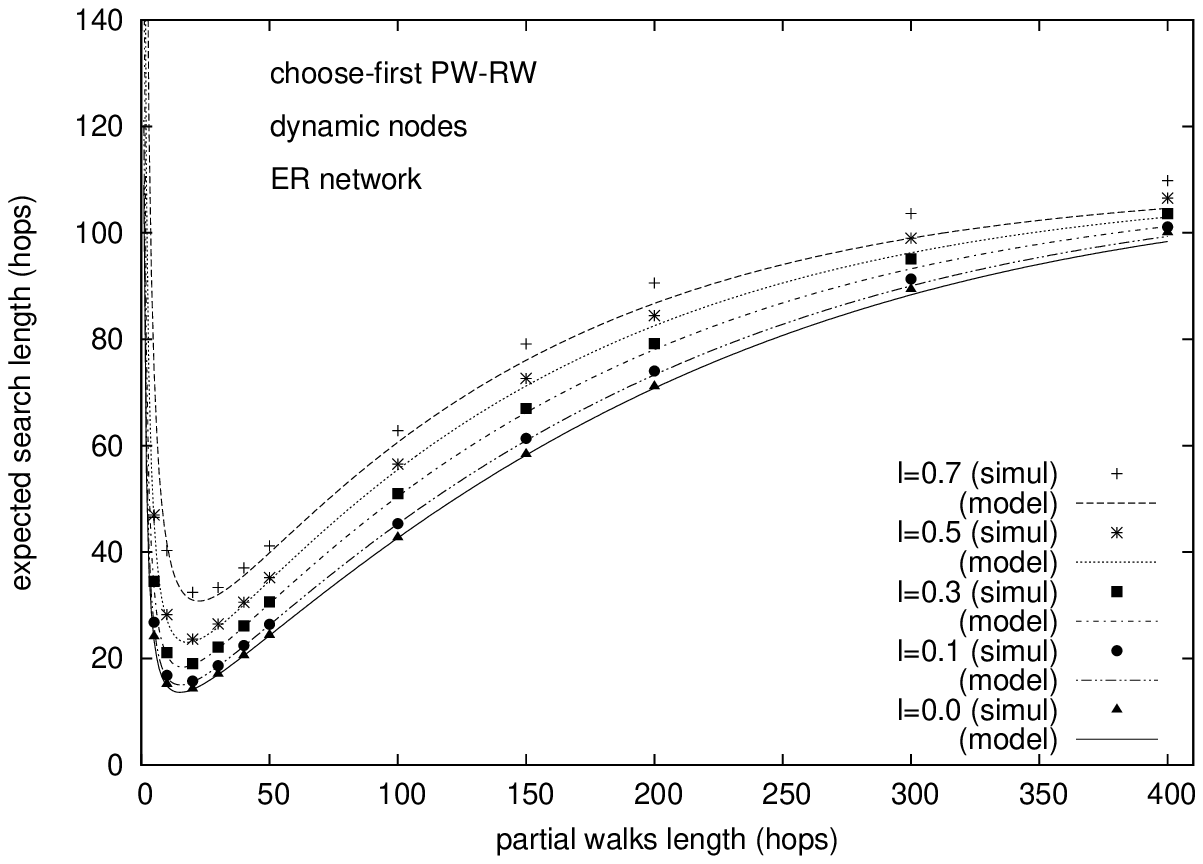}
 \caption{\emph{Choose-first PW-RW} with dynamic nodes: expected search length $\Lsexp$ vs. PW length $\s$ for several $\pd$ in a ER network.}
 \label{fig:random_pw1_dynnodes}
\end{figure}

Figures~\ref{fig:regular_pw1_dynnodes}~and~\ref{fig:random_pw1_dynnodes} show the expected search lengths achieved by \emph{choose-first PW-RW} in regular and ER networks with dynamic nodes, respectively, for several values of $l$, the probability of a node leaving the network.

\section{Check-First PW-RW with Dynamic Nodes}

\begin{figure}[h]
 \centering
 \includegraphics[width=8cm]{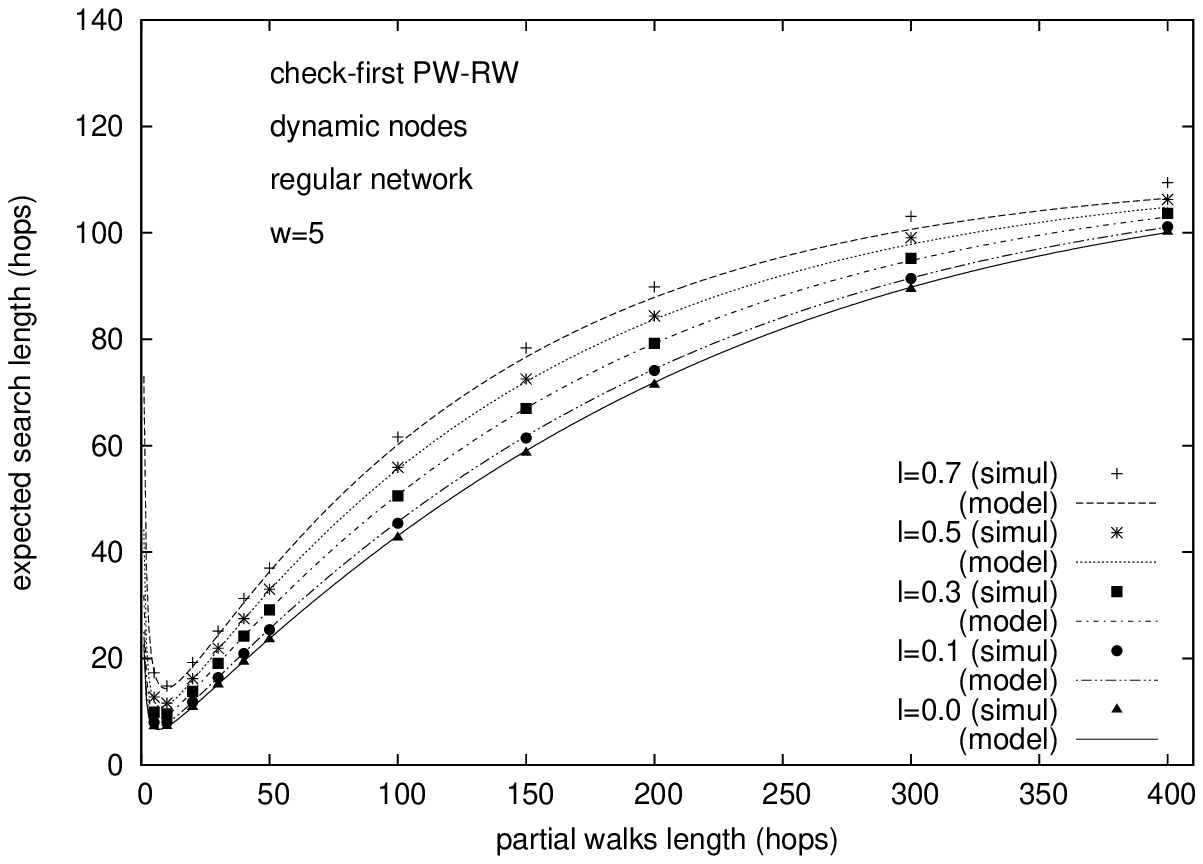}
 \caption{\emph{Check-first PW-RW} ($\w=5$) with dynamic nodes: expected search length $\Lsexp$ vs. PW length $\s$ for several $\pd$ in a regular network.}
 \label{fig:regular_pw2_dynnodes}
\end{figure}

\begin{figure}[h]
 \centering
 \includegraphics[width=8cm]{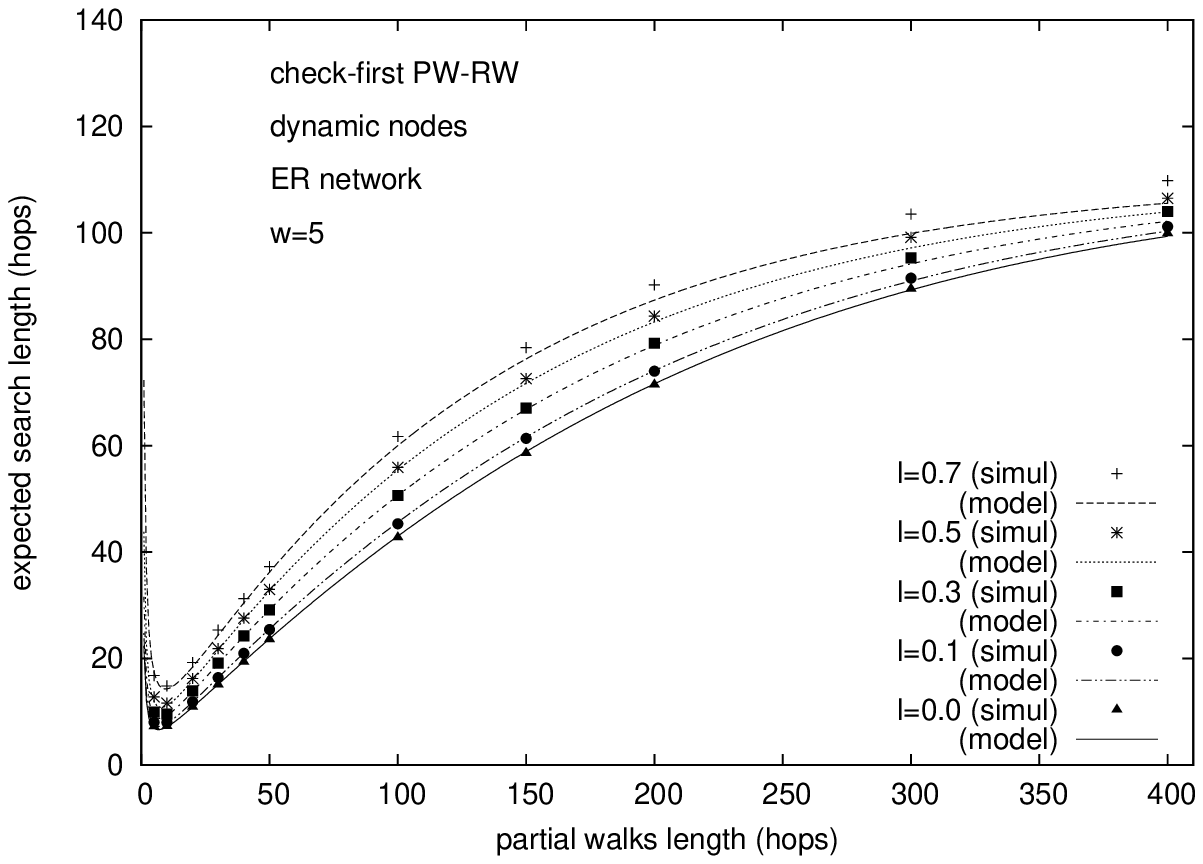}
 \caption{\emph{Check-first PW-RW} ($\w=5$) with dynamic nodes: expected search length $\Lsexp$ vs. PW length $\s$ for several $\pd$ in a ER network.}
 \label{fig:random_pw2_dynnodes}
\end{figure}

Figures~\ref{fig:regular_pw2_dynnodes}~and~\ref{fig:random_pw2_dynnodes} show the expected search lengths achieved by \emph{check-first PW-RW} in regular and ER networks with dynamic nodes, respectively, for several values of $l$.

\begin{figure}[h]
 \centering
 \includegraphics[width=8cm]{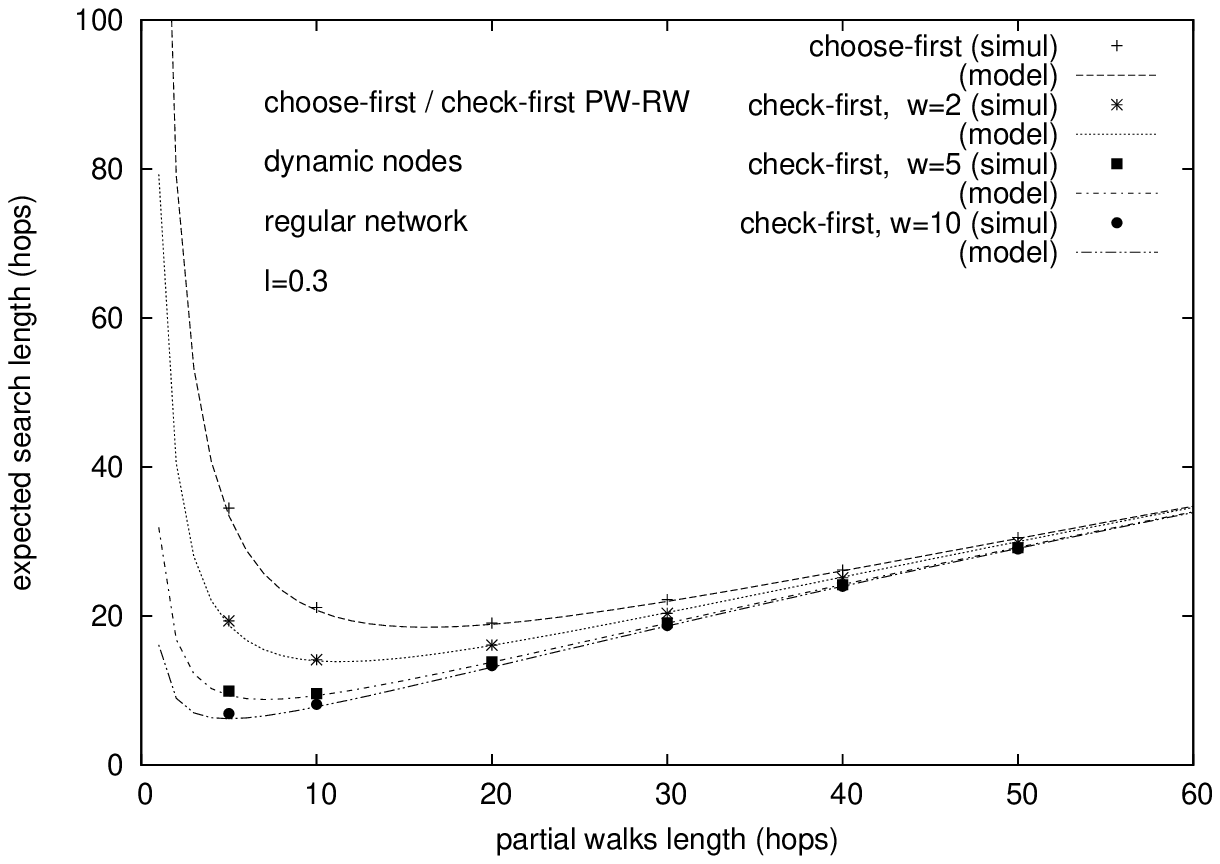}
 \caption{PW-RW with dynamic nodes: expected search length $\Lsexp$ vs. PW length $\s$ for \emph{choose-first PW-RW} and \emph{check-first PW-RW} ($\w = 2, 5, 10$) in a regular network with $\pd=0.3$.}
 \label{fig:regular_pw1_pw2_dynnodes}
\end{figure}

Finally, Figure~\ref{fig:regular_pw1_pw2_dynnodes} shows the dependency between the expected search lengths achieved by \emph{check-first PW-RW} and the number of PWs per node ($w$), in a regular network with dynamic nodes. Results for ER networks are very similar.

\end{document}